\documentclass[
	aps,
	preprint, 
	tightenlines, 
	notitlepage, 
	onecolumn, 
	preprintnumbers, 
	superscriptaddress, 
	amsmath, 
	amssymb, 
	showpacs, 
	nofootinbib]{revtex4-1}

\usepackage{amsmath}         
\usepackage{amssymb}         
\usepackage{amsfonts}        
\usepackage{graphicx}        
\usepackage{slashed}   

\usepackage{url}
\usepackage{color}
\usepackage{xcolor}
\usepackage{xspace}			
\usepackage{setspace}		

\graphicspath{{figures/}}	

\renewcommand{\tilde}{\widetilde}   

\makeatletter
\def\l@subsubsection#1#2{}
\makeatother

\newcommand{\paraflip}[1]{\underline{\textit{#1}}\xspace}

\newcommand{\PRE}[1]{{#1}}  

\newcommand{\mev}{\text{MeV}\xspace}
\newcommand{\gev}{\text{GeV}\xspace}

\newcommand{\mET}{\slashed{E}_{T}}

\renewcommand{\eqref}[1]{Eq.~(\ref{eq:#1})}
\newcommand{\eqsref}[2]{Eqs.~(\ref{eq:#1}) and (\ref{eq:#2})}
\newcommand{\secref}[1]{Sec.~\ref{sec:#1}}
\newcommand{\secsref}[2]{Secs.~\ref{sec:#1} and \ref{sec:#2}}

\newcommand{\figref}[1]{Fig.~\ref{fig:#1}}

\newcommand{\tableref}[1]{Table~\ref{table:#1}}

\newcommand{\be}{\text{Be}\xspace}
\newcommand{\bezero}{ground state $^8$Be\xspace}
\newcommand{\bestar}{$^8\be^*$\xspace}
\newcommand{\bestarprime}{$^8\be^*{}'$\xspace}
\newcommand{\Bezero}{\,^8\be}
\newcommand{\Bestar}{\,^8\be^*\xspace}
\newcommand{\Bestarprime}{\,^8\be^*{}'\xspace}
\usepackage{microtype}
\usepackage[
	colorlinks=true,
	citecolor=black,
	linkcolor=black,
	urlcolor=green!50!black,
	hypertexnames=false]{hyperref}

\begin{document}

\preprint{UCI-TR-2016-12}

\title{\PRE{\vspace*{.5in}}
Particle Physics Models for the 17 MeV Anomaly \\
in Beryllium Nuclear Decays \PRE{\vspace*{.2in}}}

\author{Jonathan L.\ Feng}
\affiliation{Department of Physics and Astronomy, University of  California, Irvine, California 92697-4575, USA
}

\author{Bartosz Fornal}
\affiliation{Department of Physics and Astronomy, University of  California, Irvine, California 92697-4575, USA
}

\author{Iftah Galon}
\affiliation{Department of Physics and Astronomy, University of  California, Irvine, California 92697-4575, USA
}

\author{Susan Gardner}
\affiliation{Department of Physics and Astronomy, University of  California, Irvine, California 92697-4575, USA
}
\affiliation{Department of Physics and Astronomy, University of 
Kentucky, Lexington, Kentucky 40506-0055, USA}

\author{Jordan Smolinsky}
\affiliation{Department of Physics and Astronomy, University of  California, Irvine, California 92697-4575, USA
}

\author{Tim M.P.\ Tait}
\affiliation{Department of Physics and Astronomy, University of  California, Irvine, California 92697-4575, USA
}

\author{Philip Tanedo\PRE{\vspace*{.2in}}}
\affiliation{Department of Physics and Astronomy, University of  California, Irvine, California 92697-4575, USA
}
\affiliation{Department of Physics and Astronomy, University of  California, Riverside, California 92521, USA \PRE{\vspace*{.5in}}
}


\begin{abstract}
\PRE{\vspace*{.2in}}
The 6.8$\sigma$ anomaly in excited $^8$Be nuclear decays via internal pair creation is fit well by a new particle interpretation. In a previous analysis, we showed that a 17~MeV protophobic gauge boson provides a particle physics explanation of the anomaly consistent with all existing constraints.  Here we begin with a review of the physics of internal pair creation in $^8$Be decays and the characteristics of the observed anomaly.  To develop its particle interpretation, we provide an effective operator analysis for excited $^8$Be decays to particles with a variety of spins and parities and show that these considerations exclude simple models with scalar particles.  We discuss the required couplings for a gauge boson to give the observed signal, highlighting the significant dependence on the precise mass of the boson and isospin mixing and breaking effects.  We present anomaly-free extensions of the Standard Model that contain protophobic gauge bosons with the desired couplings to explain the $^8$Be anomaly.  In the first model, the new force carrier is a U(1)$_B$ gauge boson that kinetically mixes with the photon; in the second model, it is a U(1)$_{B-L}$ gauge boson with a similar kinetic mixing.  In both cases, the models predict relatively large charged lepton couplings $\sim 0.001$ that can resolve the discrepancy in the muon anomalous magnetic moment and are amenable to many experimental probes.  The models also contain vectorlike leptons at the weak scale that may be accessible to near future LHC searches.
\end{abstract}

\pacs{14.70.Pw, 27.20.+n, 21.30.-x, 12.60.Cn, 13.60.-r}

\maketitle

\newpage

\begin{spacing}{1}
\tableofcontents
\end{spacing}


\section{Introduction} 

The existence of light, weakly-coupled new particles has been a well-motivated theoretical possibility for decades.  The need for dark matter has motivated these particles, either to provide the dark matter itself---for example, in the form of axions---or, more recently, to mediate interactions between the visible and dark sectors.  Grand unification provides another compelling motivation for new particles and forces.  Although these particles and forces are typically expected to be heavy and short-range, respectively, it is possible that a remnant of grand unification might survive down to low energies.   Another independent, but related, possibility is that some linear combination of the ``accidental'' $B$ and $L_i$ global symmetries of the Standard Model (SM) might be gauged.  If these symmetries are spontaneously broken at low energies, they must also be weakly-coupled.  All of these provide ample motivation for a diverse program of high-statistics searches for new particles far from the energy frontier.

Nuclear transitions provide a means to probe light, weakly-coupled new physics.  Indeed, in 1978, Treiman and Wilczek~\cite{Treiman:1978ge}, as well as Donnelly et al.~\cite{Donnelly:1978ty}, proposed that axions could be discovered through the study of nuclear decays. Such searches are now  established as part of the corpus of constraints on axions and axion-like particles~\cite{Savage:1986ty, Savage:1988rg}, as well as on light scalar particles with Higgs-like couplings~\cite{Freedman:1984sd}. The possible new particles include not only scalars and pseudoscalars, but also those with other spin-parity assignments, which may manifest themselves in different nuclear transitions.  There are many possible nuclear transitions to study, but particularly promising are those that can be studied through excited nuclear states that are resonantly produced in extraordinary numbers, providing a high-statistics laboratory to search for \mev-scale new physics.  

Krasznahorkay et al.\ have recently observed unexpected bumps in both the distributions of opening angles and invariant masses of electron--positron pairs produced in the decays of an excited $^8$Be nucleus~\cite{Krasznahorkay:2015iga}. The bump in the angular distribution appears against monotonically decreasing backgrounds from SM internal pair creation (IPC), and the anomaly has a high statistical significance of 6.8$\sigma$. The shape of the excess is remarkably consistent with that expected if a new particle is being produced in these decays, with the best fit to the new particle interpretation having a $\chi^2$-per-degree-of-freedom of 1.07.  

In previous work~\cite{Feng:2016jff}, we examined possible particle physics interpretations of the $^8$Be signal.  We showed that scalar and pseudoscalar explanations are strongly disfavored, given  mild assumptions.  Dark photons $A'$, massive gauge bosons with couplings to SM particles that are proportional to their electric charge~\cite{Kobzarev:1966qya, Okun:1982xi, Holdom:1985ag, Holdom:1986eq}, also cannot account for the $^8$Be anomaly, given constraints from other experiments. The most stringent of these is null results from searches for $\pi^0 \to A' \gamma$. This can be circumvented if the new spin-1 state couples to quarks vectorially with suppressed couplings to the proton. We concluded that a new ``protophobic'' spin-1 boson $X$, with mass around 17~MeV and mediating a weak force with range 12~fm, provides an explanation of the $^8$Be anomaly consistent with all existing experimental constraints. The implications and origins of a protophobic gauge boson have been further studied in Refs.~\cite{Gu:2016ege,Chen:2016dhm,Liang:2016ffe}.

Protophobic gauge bosons are not particularly unusual.  The $Z$ boson is  protophobic at low energies, as is any new boson that couples to $B-Q$, the difference of the baryon-number and electric currents.   As we show below, it is extremely easy to extend the SM to accommodate a light gauge boson with protophobic quark couplings.  Simultaneously satisfying the requirements on lepton couplings requires more care.  To produce the observed $e^+e^-$ events, the coupling to electrons must be non-zero. This coupling is bounded from above by the shift the new boson would induce on the electron magnetic dipole moment and from below by searches for dark photons at beam dump experiments. The neutrino coupling, in turn, is bounded by $\nu$--$e$ scattering experiments, as well as by the non-observation of coherent neutrino--nucleus scattering. Any model that consistently explains the $^8$Be signal must satisfy all of these constraints.  

In \secref{beryllium} we review the $^8$Be system and the observed anomaly.  In \secref{effective:theory}, we present an effective operator analysis of the $^8$Be nuclear transitions and consider a variety of spin-parity assignments for the new boson. We show that many simple candidates, including scalars and pseudoscalars, are excluded, while a protophobic spin-1 gauge boson is a viable candidate. In the next three sections, we consider in detail the couplings such a gauge boson must have to explain the $^8$Be anomaly: in \secref{nuclear:physics:effects} we discuss the impact of isospin mixing and breaking in the $^8$Be system; in \secref{experimental:effects} we discuss the required gauge boson couplings to explain the signal, noting the sensitivity to the gauge boson's precise mass; and in \secref{constraints} we evaluate the constraints imposed by all other experiments, refining the discussion in Ref.~\cite{Feng:2016jff}, especially for the neutrino constraints.  With this background, in \secsref{modelB}{modelB-L}, we construct simple, anomaly-free extensions of the SM that contain protophobic gauge bosons with the desired couplings to explain the $^8$Be anomaly.  In \secref{future:experiments} we discuss current and near-term experiments that may test this new particle explanation, and we conclude in \secref{conclusions} by summarizing our results and noting some interesting future directions.


\section{The $^8$Be Anomaly}
\label{sec:beryllium}

\subsection{$^8$Be Spectrum and Electromagnetic Decays}
\label{sec:spectrum}

\begin{figure}[t]
\includegraphics[width=.95\linewidth]{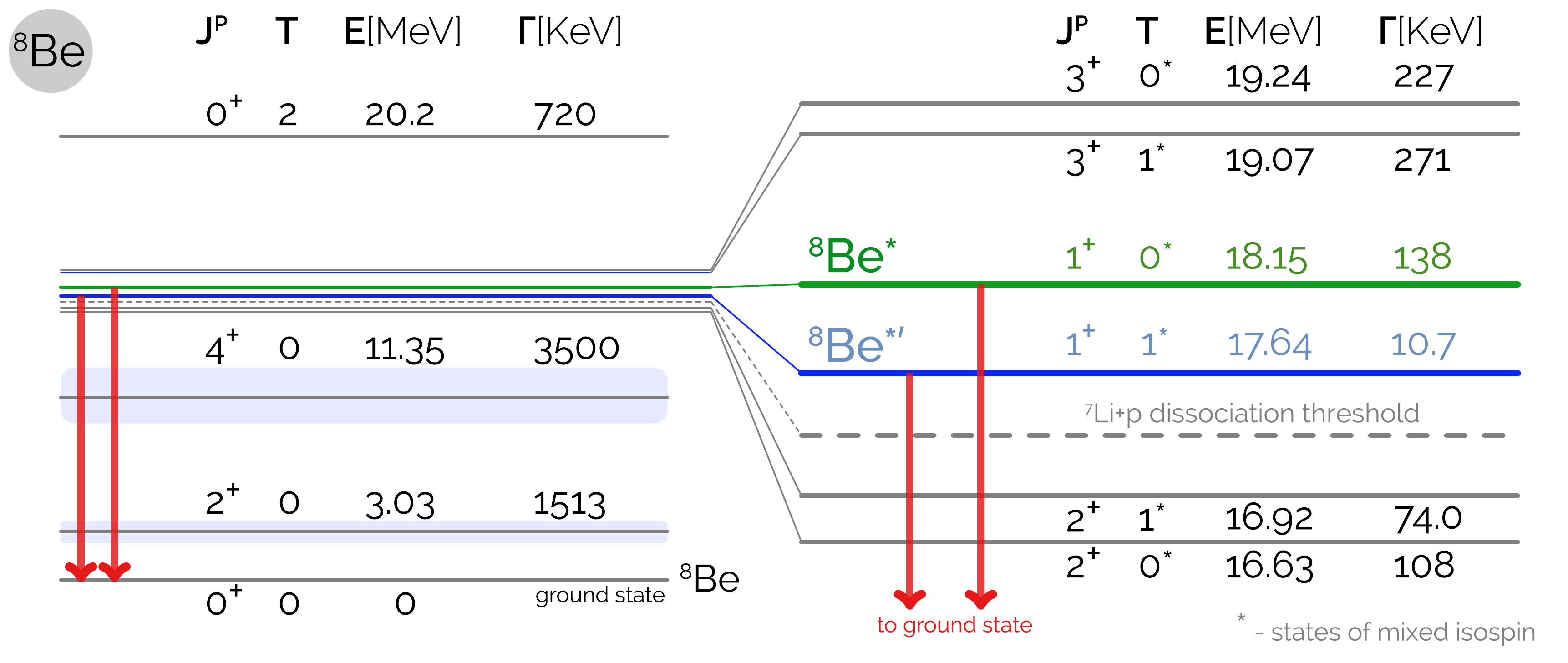} 
\vspace*{-0.1in}
\caption{The most relevant $^8$Be states, our naming conventions for them, and their spin-parities $J^P$, isospins $T$, excitation energies $E$, and decay widths $\Gamma$ from Ref.~\cite{nndc}. Asterisks on isospin assignments indicate states with significant isospin mixing.  Decays of the \bestar (18.15) state to the \bezero exhibit anomalous internal pair creation; decays of the \bestarprime (17.64) state do not~\cite{Krasznahorkay:2015iga}. 
\label{fig:levels}}
\end{figure}

We review relevant properties of the $^8$Be system.  Some of the energy levels of $^8$Be are shown in Fig.~\ref{fig:levels}.  The ground state of the $^8$Be nucleus is only 0.1~\mev above the threshold for $\alpha \alpha$ breakup, and $\alpha$ clustering is thought to inform its structure and excitations~\cite{BohrMottelson,Datar:2013pbd}. The ground state is a spin-parity $J^P= 0^+$ state with isospin $T=0$, and its lowest-lying excitations are $2^+$ and $4^+$ rotational states, nominally of its $\alpha\alpha$ dumbbell-shape, with $T=0$, excitation energies 3.03 MeV and 11.35 MeV, and decay widths 1.5~\mev and 3.5~\mev, respectively.\footnote{The $2^+$ excited state is notable in that it is produced through the $\beta$ decay of $^8$B~\cite{Bahcall:1996qv}. It is pertinent to the solar neutrino problem because it appears with a neutrino of up to 14 MeV in energy, and it also enters in precision tests of the symmetries of the charged, weak current in the mass eight system~\cite{Sternberg:2015nnr}.}

Going up in excitation energy, the next lowest lying states are isospin doublets of $T=0,1$ states with spin-parity assignments of $2^+$, $1^+$, and $3^+$, respectively.  The $2^+$ states are of such mass that the $\alpha\alpha$ final state is the only particle-decay channel open to them. Since both states are observed to decay to the $\alpha\alpha$ final state~\cite{nndc}, which has $T=0$, the $2^+$ states are each regarded as mixtures of the $T=0$ and $T=1$ states. The qualitative evidence for isospin-mixing in the $1^+$ and $3^+$ states is less conclusive, but each doublet state is regarded as a mixed $T=0$ and $T=1$ state~\cite{Barker:1966}. 

In this paper, our focus is on the transitions of the $1^+$ isospin doublet to the ground state, as illustrated in Fig.~\ref{fig:levels}.  We refer to the ground state as simply $^8$Be and to the $1^+$ excited states with excitation energies 18.15 MeV and 17.64 MeV as \bestar and \bestarprime, respectively .  As noted above, these latter two states mix, but \bestar is predominantly $T=0$, and \bestarprime is predominantly $T=1$. The properties of these states and their electromagnetic transitions have been analyzed using quantum Monte Carlo (QMC) techniques using realistic, microscopic Hamiltonians~\cite{Wiringa:2000gb,Pieper:2004qw,Wiringa:2013fia,Pastore:2014oda}.  We discuss the current status of this work and its implications for the properties of new particles that may be produced in these decays in \secref{nuclear:physics:effects}. 

The particular transitions relevant for the observed anomaly are internal pair conversion (IPC) decays. This is a process in which an excited nucleus decays into a lower-energy state through the emission of an electron--positron pair~\cite{deSandF,Schluter1981327, siegbahn2012alpha}.  Like $\gamma$-decays---which satisfy selection rules based on angular momentum and parity---these decays can be classified by their parity (electric, E, or magnetic, M) and partial wave $\ell$. A $p$-wave magnetic transition, for example, is labeled M1. The spectra of electron--positron invariant masses and opening angles in these decays are known to be monotonically decreasing for each partial wave in the SM~\cite{Rose:1949zz}. It is customary to normalize the IPC rate with respect to that of $\gamma$ emission for the same nuclear transition, when the latter exists. This is because the nuclear matrix elements, up to Coulomb corrections, as well as some experimental systematic errors, cancel in this ratio.  $^8$Be, moreover, is of sufficiently low-$Z$ that the effects of its Coulomb field on IPC are negligible~\cite{Schluter1981327}. \bestar decays to $^7$Li $p$ most of the time, but its electromagnetic transitions have branching fractions $\text{Br}(\Bestar \to {}^8\text{Be} \, \gamma) \approx 1.4 \times 10^{-5}$~\cite{Tilley:2004zz} and $\text{Br}(\Bestar \to {}^8\text{Be} \; e^+ e^-) \approx 3.9\times 10^{-3} \, \text{Br}(\Bestar \to {}^8\text{Be} \, \gamma)$~\cite{Rose:1949zz, Schluter1981327}.

\subsection{The Atomki Result}

\begin{figure}[t]
\includegraphics[width=.75\linewidth]{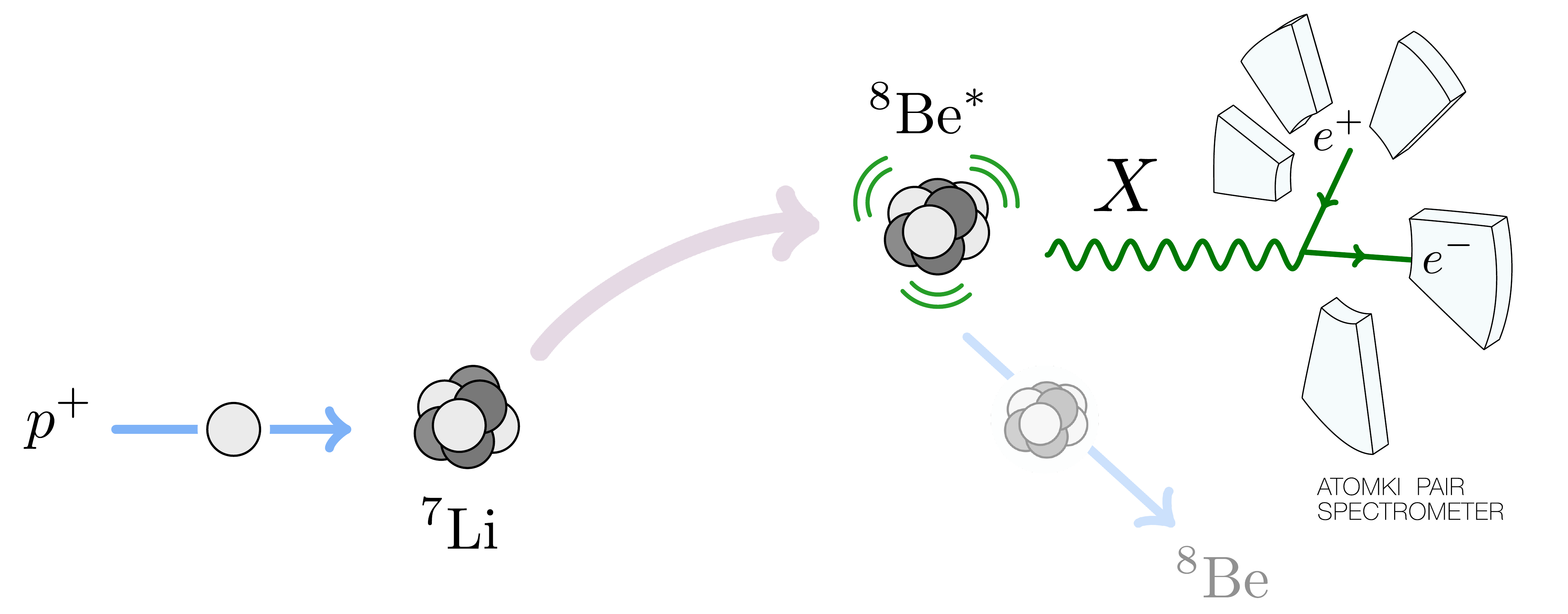} 
\vspace*{-0.1in}
\caption{Schematic depiction of the Atomki pair spectrometer experiment~\cite{Gulyas:2015mia, Krasznahorkay:2015iga}, interpreted as evidence for the production of a new boson $X$. The proton beam's energy is tuned to excite lithium nuclei into the \bestar state, which subsequently decays into the $^8$Be ground state and $X$. The latter decays into an electron--positron pair whose opening angle and invariant mass are measured. 
\label{fig:schematic} }
\end{figure}

The Atomki pair spectrometer has observed the IPC decays of $\Bestar$ with high statistics~\cite{Gulyas:2015mia, Krasznahorkay:2015iga}.  A sketch of the experiment and the new physics process being probed is shown in \figref{schematic}. A beam of protons with kinetic energies tuned to the resonance energy of 1.03 MeV collide with Li nuclei to form the resonant state \bestar, and a small fraction of these decay via $\Bestar \to {}^8\text{Be}\, e^+ e^-$.  The spectrometer is instrumented with plastic scintillators and multi-wire proportional chambers in the plane perpendicular to the proton beam. These measure the electron and positron energies, as well as the opening angle of the $e^+e^-$ pairs that traverse the detector plane, to determine the distributions of opening angle $\theta$ and invariant mass $m_{ee}$.  

The experiment does not observe the SM behavior where the $\theta$ and $m_{ee}$ distributions fall monotonically. Instead, the $\theta$ distribution exhibits a high-statistics bump that peaks at $\theta \approx 140^\circ$ before returning to near the SM prediction at $\theta \approx 170^\circ$~\cite{Krasznahorkay:2015iga}.  To fit this distribution, Krasznahorkay~et al.\ consider many possible sources, including the M1 component from IPC, but also others, such as an E1 component from non-resonant direct proton capture~\cite{Fisher:1976zza}. They observe that the best fit comes from a 23\% admixture of this E1 component.  Nevertheless, they are unable to explain the bump by experimental or nuclear physics effects, and instead find that the excess in the $\theta$ distribution has a statistical significance of 6.8$\sigma$~\cite{Krasznahorkay:2015iga}.  A corresponding bump is seen in the $m_{ee}$ distribution.  

If a massive particle is produced with low velocity in the \bestar decay and then decays to $e^+ e^-$ pairs, it will produce a bump at large opening angles.  It is therefore natural to consider a new particle $X$ and the two-step decay $\Bestar \to {}^8\text{Be}\, X$ followed by $X \to e^+ e^-$.  With fixed background, Krasznahorkay~et al.\ find that the best fit mass and branching fraction are~\cite{Krasznahorkay:2015iga}
\begin{eqnarray}
&&m_X = 16.7 \pm 0.35 \,(\text{stat}) \pm 0.5\,(\text{sys})~\mev  \label{eq:Krasznahorkay:result1} \\
&&	\frac{\Gamma(\Bestar \to \Bezero\, X)}{\Gamma(\Bestar \to \Bezero\, \gamma)} \,
	\text{Br}(X\to e^+e^-) = 5.8 \times 10^{-6} \ .
	\label{eq:Krasznahorkay:result}
\end{eqnarray}
For the best fit parameters, the fit to this new particle interpretation is excellent, with a $\chi^2/\text{dof} = 1.07$.

The new particle interpretation passes a number of simple consistency checks.  The electron--positron invariant mass and opening angle are related by 
\begin{eqnarray}
m_{ee}^2 &=& 2 E_{e^+} E_{e^-} - 2 \sqrt{E_{e^+}^2 - m_e^2} 
\sqrt{E_{e^-}^2 - m_e^2} \cos\theta + 2m_e^2 \nonumber \\
&=& (1-y^2) E^2 \sin^2 \frac{\theta}{2} + 
2m_e^2\left(1+\frac{1+y^2}{1-y^2}\cos\theta \right) +\mathcal O(m_e^4)  \ ,
\label{eq:mee}
\end{eqnarray}
where 
\begin{equation}
	E \equiv E_{e^+} + E_{e^-} \qquad \text{and} \qquad
	y \equiv \frac{E_{e^+} - E_{e^-}}{E_{e^+} + E_{e^-}} 
	\label{eq:energy:asymmetry:variables}
\end{equation}
are the total energy and energy asymmetry, respectively.  The second term in the last line of \eqref{mee} is much smaller than the first and may be neglected. At the Atomki pair spectrometer, the $\Bestar$ nuclei are produced highly non-relativistically, with velocity of $0.017c$ and, given $m_X \approx 17~\mev$, the $X$ particles are also not very relativistic.  As a result, the $e^+$ and $e^-$ are produced with similar energies, and so one expects small $|y|$ and $m_{ee} \approx E \sin (\theta/2)$.  The excesses in the $\theta$ and $m_{ee}$ distributions satisfy this relation.

The Atomki collaboration verified that the excess exclusively populates the subset of events with $|y| \leq 0.5$ and is absent in the complementary $|y| > 0.5$ domain, that the excess appears and then disappears as one scans through the proton beam resonance kinetic energy of 1.03 MeV, and that the excess becomes more pronounced when restricting to the subset of events with $E > 18~\mev$ and is absent for lower energy events~\cite{Krasznahorkay:2015iga}.  The latter two observations strongly suggest that the observed IPC events are indeed from $\Bestar$ decays rather than 
from interference effects and that the decays go to the ground state $^8$Be, as opposed to, for example, the broad 3~\mev $J^P = 2^+$ state.  Decays to the 3~\mev state would have a maximum total energy of 15~\mev and do not pass the $E > 18~\mev$ cut even when including effects of the energy resolution, which has a long low-energy tail, but not a high-energy one (see Fig.~2 of Ref.~\cite{Gulyas:2015mia}).\footnote{The widths of the $m_{ee}$ and $\theta$ distributions are determined by the $\mathcal O(\mev)$ energy resolution for the electrons and positrons~\cite{Gulyas:2015mia}, which should not be confused with the 10 keV energy resolution for $\gamma$-rays used in testing the target thickness~\cite{Krasznahorkay:2015iga}.}

Finally, we note that IPC decays of the 17.64~\mev, isotriplet \bestarprime state have also been investigated at the Atomki pair spectrometer.  An anomaly had previously been reported in \bestarprime decays~\cite{deBoer1996235}.  This anomaly was featureless and far easier to fit to background than the bumps discussed here, and it has now been excluded by the present Atomki collaboration~\cite{Gulyas:2015mia}. If the observed anomaly in \bestar decays originates from a new particle, then the absence of new particle creation in the \bestarprime decay combined with the isospin mixing discussed in \secref{nuclear:physics:effects} strongly suggest that such decays are kinematically---not dynamically---suppressed and that the new particle mass is in the upper part of the range given in \eqref{Krasznahorkay:result1}.  It also suggests that with more data, a similar, but more phase space-suppressed, excess may appear in the IPC decays of the 17.64 state.

\section{Nuclear Effective Theory and Particle Candidates}
\label{sec:effective:theory}

The transition $\Bestar \to \Bezero \, X$ followed by $X \to e^+ e^-$ implies that $X$ is a boson.  We consider the cases in which it is a scalar or vector particle with positive or negative parity.  In this reaction, its de Broglie wavelength is $\lambda \sim (6~\mev)^{-1}$, much longer than the characteristic size of the $^8$Be nucleus, $r \sim (100~\mev)^{-1}$.  In this regime, the nucleus looks effectively point-like, and one can organize the corrections from the nuclear structure as a series in $r / \lambda$.  This approach has a long and fruitful history in the analysis of radiative corrections in weak nuclear decays~\cite{Sirlin:1967zza, petrov2015effective}.

We perform such an analysis for the case of \bestar decaying to a new boson $X$. Many theories predict the existence of new, weakly-coupled, light degrees of freedom that, prima facie, may play the role of the $X$ boson. We show that some common candidates for $X$ are excluded. 
We note that for the case where $X$ has spin-parity $J^P = 1^-$ and isospin mixing between \bestar and \bestarprime is neglected,  nuclear matrix elements and their uncertainties cancel in the ratio of partial widths, $\Gamma(\Bestar \to \Bezero \, X) / \Gamma(\Bestar \to \Bezero \, \gamma)$.

\subsection{Effective Operators for $\Bestar \to \Bezero \, X$}

The candidate spin/parity choices for $X$ are: a $0^-$ pseudoscalar $a$, a $1^+$ axial vector $A$, and a $1^-$ vector $V$. We argue below that there is no scalar operator in the parity-conserving limit. The leading Lorentz- and parity-invariant operators mediating the transition $\Bestar \to \Bezero \, X$ are: 
\begin{eqnarray}
	\mathcal L_P &=& g_P \Bezero \,  \left(\partial_\mu a\right) {}^8\text{Be}^{*\mu} \\
	\mathcal L_A &=&  \frac{g_A}{\Lambda_A} \Bezero \,  G^{\mu\nu} F^{(A)}_{\mu\nu}
	+ \frac{g'_A}{\Lambda_A} m_A^2 \Bezero\, A_\mu \, {}^8\text{Be}^{*\mu}  \label{eq:axial} \\
	\mathcal L_V &=& \frac{g_V}{\Lambda_V}\Bezero \, G_{\mu\nu} F^{(V)}_{\rho\sigma} 
		\epsilon^{\mu\nu\rho\sigma} \label{eq:vector} \ , 
\end{eqnarray}
where $G_{\mu\nu} \equiv \partial_\mu \Bestar_\nu - \partial_\nu \Bestar_\mu$ is the field strength for the excited \bestar state, $F^{(V)}_{\mu\nu}$ and $F^{(A)}_{\mu\nu}$ are the field strengths for the new vector and axial vector bosons, respectively, and the dimensionful parameters $\Lambda_i$ encode the dominant nuclear matrix elements relevant for the transition in each case. 

In the vector case, Lorentz and parity invariance requires that all operators containing the fields $^8\text{Be}$, $^8\text{Be}_\mu^*$, and $V_\mu$ must also contain two derivatives and $\epsilon_{\mu \nu \rho \sigma}$. Any operators in which the two derivatives act upon the same field vanish under antisymmetrization of the Lorentz indices, so that the only other possible operators are

\begin{equation}
(\partial_\mu \Bezero ) \, \Bestar_\nu \, F_{\rho \sigma}^{(V)} \epsilon^{\mu \nu \rho \sigma} \qquad \text{and} \qquad ( \partial_\mu \Bezero ) \, G_{\nu \rho} V_\sigma \epsilon^{\mu \nu \rho \sigma} \ .
\end{equation}

However, these operators can each be integrated by parts to produce a term that vanishes by antisymmetrization, and the unique operator in $\mathcal L_V$. This is in contrast to the axial vector case, where the gauge-breaking part cannot be related by operator identities to the gauge-invariant piece and is thus a separate term with a separate effective coupling $g'_A$.

\subsection{Scalar Candidates}

A popular example of a $J^P = 0^+$ scalar candidate for the $X$ boson is a dark Higgs~\cite{Batell:2009yf}. However, a scalar cannot mediate the observed \bestar decay in the limit of conserved parity. The initial \bestar state has unit angular momentum and is parity-even, $J^P = 1^+$.  Angular momentum conservation requires the final state $\Bezero \, X$, which consists of two $0^+$ states, to have orbital angular momentum $L = 1$. This, however, makes the final state parity-odd while the initial state is parity-even. This implies that there are no Lagrangian terms in a parity-conserving effective field theory that couple a scalar to the \bestar and \bestarprime. This can also be seen at an operator level; for example, the operator $(\partial_\mu S) (\partial_\nu \Bezero) G_{\rho\sigma} \epsilon^{\mu\nu\rho\sigma}$ vanishes upon integrating by parts and using the Bianchi identity.

\subsection{Pseudoscalar Candidates}

A $J^P = 0^-$ pseudoscalar or axion-like particle, $a$, generically has a coupling to photons of the form $g_{a \gamma \gamma} a F^{\mu\nu} \tilde{F}_{\mu\nu}$ that is generated by loops of charged particles~\cite{Weinberg:1977ma,Wilczek:1977pj,Preskill:1982cy}. For a mass $m_a \approx 17~\mev$, all values of this coupling in the range $(10^{18}~\gev)^{-1} < g_{a \gamma \gamma} < (10~\gev)^{-1}$ are experimentally excluded~\cite{Hewett:2012ns,Dobrich:2015jyk}. These bounds may be significantly revised in the presence of non-photonic couplings, however.

\subsection{Axial Vector Candidates}

Axial vector candidates have several virtues.  First, as we show in \secref{constraints}, one of the most restrictive constraints on the $X$ particle comes from the decay of neutral pions, $\pi^0\to X\gamma$. If $X$ is an axial vector, this decay receives no contribution from the axial anomaly and non-anomalous contributions to pion decay vanish in the chiral limit by the Sutherland--Veltman theorem~\cite{Sutherland:1967vf,Veltman}.  

Second, unlike the other spin--parity combinations, the axial candidate has two leading-order effective operators with different scaling with respect to the $X$ three-momentum. The $g_A'$ term in \eqref{axial} yields a \bestar decay rate that scales as $\Gamma_X \sim |\mathbf{k}_X|$, whereas the $g_{A,V}$ terms induce rates that scale as $\Gamma_X \sim |\mathbf{k}_X|^3$. Thus, the axial particle may produce the observed anomalous IPC events with smaller couplings than the vector, $g_A' \ll g_{V}$. This may help avoid some of the other experimental bounds discussed in \secref{constraints}.  At the same time, there may be more severe constraints, as we discuss in \secref{conclusions}.

Unfortunately, large uncertainties in the nuclear matrix element for axial vectors make it difficult to extract the required couplings for this scenario. To the best of our knowledge, there is no reliable \textit{ab initio} calculation or measurement of the matrix element we would need
in the $^8$Be system.

\subsection{Vector Candidates}
\label{sec:vector}

The primary candidate for the $X$ boson and the focus for the remainder of this study is a $J^P = 1^-$ vector. A new vector couples to a current $J_X^\mu$ that is a linear combination of the SM fermion currents,
\begin{align}
	\mathcal L &\supset i X_\mu J_X^\mu 
	= i X_\mu \sum_{i=u,d,\ell,\nu} \varepsilon_i e J_i^\mu\, ,
	&
	J_i^\mu = \bar f_i \gamma^\mu f_i \ .
	\label{eq:vector:current:and:couplings}
\end{align}
Here we have introduced separate couplings to up-type quarks, down-type quarks, charged leptons, and neutrinos and assigned them charges $\varepsilon_i$ in units of $e$.  Family-universal couplings of this type naturally avoid the introduction of tree-level flavor-changing neutral currents, which are highly constrained.  Conservation of $X$ charge implies that the couplings to the proton and neutron currents, $J_p^\mu$ and $J_n^\mu$, are determined by $\varepsilon_p = 2\varepsilon_u + \varepsilon_d$ and $\varepsilon_n = \varepsilon_u + 2 \varepsilon_d$, so that 
\begin{align}
	J_X^\mu = 
	\sum_{i=u,d,\ell,\nu} \varepsilon_i e J_i^\mu
	= (2\varepsilon_u+\varepsilon_d)eJ_p^\mu 
	+ (\varepsilon_u+2\varepsilon_d)J_n^\mu 
	+ \varepsilon_\ell e J_\ell^\mu
	+ \varepsilon_\nu e J_\nu^\mu \ .
\end{align}
For the low energies at which we work, it is important to map this quark-level expression to 
one in terms of hadrons.  Denoting the current in the previous
equation by $J_X^{\mu\, ({\rm quark})}$, we effect this by matching the requisite 
matrix element to its equivalent in hadronic degrees of freedom. That is, for the proton, 
\begin{align}
J_p^\mu \equiv 
\langle p(p') | J_X^{\mu \, ({\rm quark})} | p(p) \rangle  & =
e \overline{u}_p(p')  \left\{ F^X_{1,p} (q^2)  
\gamma^\mu + F^X_{2,p} (q^2)
\sigma^{\mu \nu} q_\nu / 2 M_p \right\} u_p (p) \,,  
\end{align}
where $| p(p) \rangle$ denotes a proton state composed of quarks and $u_p (p)$ is the 
Dirac spinor of a free proton. Note that QCD generates all the possible currents
compatible with Lorentz invariance and electromagnetic current conservation. We choose 
$F^X_{1,p} (q^2)$ and  $F^X_{2,p} (q^2)$ to denote the X-analogues of the familiar Dirac
and Pauli form factors. Finally we form the analogue of the Sachs magnetic form factor 
by introducing $G_{M,p}^X(q^2) =  F^X_{1,p} (q^2) + F^X_{2,p} (q^2)$, recalling that 
$G_{M,p}(0)$ is given by the total magnetic moment of the proton 
--- we refer to Ref.~\cite{Perdrisat:2006hj} for a review. 
The M1 transition of interest here is determined by the total magnetic moment operator. 

The nucleon currents, written in either the quark or hadron basis, 
can, in turn, be combined to form isospin currents 
\begin{align}
	J_{0}^\mu &= J_p^\mu + J_n^\mu
	& 
	J_{1}^\mu &= J_p^\mu - J_n^\mu
	\ . 
	\label{eq:X:isospin:current}
\end{align}
Assuming isospin is conserved and the $^8$Be states are isospin eigenstates, $\langle \Bezero | J_1^\mu |\Bestar\rangle = 0$, since both \bestar and the \bezero are isosinglets. In this case, 
\begin{align}
	\langle \Bezero | J_X^\mu |\Bestar\rangle 
	&= \frac{e}{2} (\varepsilon_p + \varepsilon_n)
	\langle \Bezero | J_0^\mu |\Bestar\rangle 
	\label{eq:isospin:matrix:elements1}
	\\
	\langle \Bezero | J_\text{EM}^\mu |\Bestar\rangle 
	&= \frac{e}{2} 
	\langle \Bezero | J_0^\mu |\Bestar\rangle \ .
	\label{eq:isospin:matrix:elements}
\end{align}
The $J_0$ nuclear matrix elements therefore cancel in the ratio $\Gamma(\Bestar \to \Bezero\, X) / \Gamma(\Bestar \to \Bezero\, \gamma)$. This observation may be modified significantly when isospin violation is included, as we discuss in \secref{nuclear:physics:effects}.

If one sets $g_V = e$ and identifies $F^{(V)}_{\rho\sigma}$ with the electromagnetic field strength in \eqref{vector}, then the leading operator in $\mathcal L_V$  describes the ordinary electromagnetic transition via $\gamma$ emission. Indeed, in this SM case, Lorentz- and parity-invariance require the characteristic $\Gamma(\Bestar \to \Bezero\, \gamma) \propto |\mathbf k_\gamma|^3$ momentum dependence of an M1 transition. The matrix elements in \eqsref{isospin:matrix:elements1}{isospin:matrix:elements} thus imply that $\Lambda_V$ in \eqref{vector} is universal for spin-1 particles.  Combining all of these pieces, we find that  
\begin{align}
	\frac{
	\Gamma(\Bestar \to \Bezero\, X)
	}{
	\Gamma(\Bestar \to \Bezero\, \gamma)
	}
	&= 
	(\varepsilon_p+\varepsilon_n)^2
	\frac{|\mathbf{k}_X|^3}{|\mathbf{k}_\gamma|^3}
	=
	(\varepsilon_p+\varepsilon_n)^2
	\left[1 - \left(\frac{m_X}{18.15\text{ MeV}}\right)^2\right]^{3/2}
	\, ,
	\label{eq:IPCC:for:X}
\end{align}
when isospin is conserved. This is a convenient expression, as the experimental best fit for the anomalous decay rate to a new vector $X$ is presented in terms of this ratio of decay widths, as seen in \eqref{Krasznahorkay:result}. 

A simple, well-known vector boson candidate is the dark photon $A'$~\cite{Kobzarev:1966qya, Okun:1982xi, Holdom:1985ag, Holdom:1986eq}. The dark photon is a light particle that can have small, but technically natural, couplings to the SM. For a given mass, the dark photon interactions are controlled by a single kinetic mixing parameter, $\varepsilon$.  This is related to the effective coupling in \eqref{vector} by $g_V = \varepsilon e$. Substituting this into \eqref{IPCC:for:X} and comparing to the experimental result in \eqref{Krasznahorkay:result}, one finds that $\varepsilon^2 \approx 10^{-4}$, which is experimentally excluded by, for example, $\pi^0\to A' \gamma$ searches at NA48/2~\cite{Batley:2015lha}.\footnote{Ref.~\cite{Krasznahorkay:2015iga} quotes a fit of $\varepsilon^2 \sim 10^{-7}$. The discrepancy appears to come from the use of expressions for axions~\cite{Donnelly:1978ty} rather than dark photons.} 

A generalization of the dark photon idea is to consider also mixing between the new boson and the SM $Z$. Such a particle is spin-1 with no definite parity. Unfortunately, bounds from atomic parity violation are extremely stringent~\cite{Davoudiasl:2012qa} and constrain the dark $Z$ couplings to be too small to explain the $^8$Be anomaly.

Another type of spin-1 particle is a light baryon-minus-lepton number ($B-L$)  boson~\cite{Mohapatra:1980qe,Buchmuller:1991ce,Buchmuller:1992qc}. This scenario is constrained by neutrino scattering off electrons and, assuming no kinetic mixing, provides the upper limit $g_{B-L} \lesssim  2 \times 10^{-5}$~\cite{Bilmis:2015lja}, which is again too small to account for the excess.  

As we discuss in detail in \secref{nuclear:physics:effects}, \eqref{IPCC:for:X} may receive significant corrections in the presence of isospin mixing and breaking.  We will also see, however, that in the experimentally viable limit of $\varepsilon_p \ll \varepsilon_n$, these corrections are small. For the cases of the dark photon, dark $Z$, and $B-L$ gauge boson discussed above, the size of the $^8$Be signal and the strength of the constraints on $\pi^0 \to X \gamma$ essentially enforce protophobia, and so the arguments against these candidates remain.

\section{Signal Dependence on Isospin Mixing and Breaking} 
\label{sec:nuclear:physics:effects}

The discussion of \secref{vector} assumed that isospin is conserved and that the $^8$Be states are states of well-defined isospin. As noted in \secref{spectrum}, however, there is substantial evidence that the $^8$Be states are isospin-mixed, and, as we note below, there may also be isospin breaking in the electromagnetic transition operators stemming from the neutron--proton mass difference. 
In this section, we determine the impact of isospin mixing and breaking on the rate for $\Bestar \to \Bezero\, X$, which, of course, has implications for the parton-level couplings required to explain the $^8$Be signal.

The ground-state structure and excitation spectrum of $^8$Be, 
as well as its electromagnetic transitions, 
have been studied with {\em ab initio} QMC techniques, based on 
non-relativistic Hamiltonians with phenomenological nucleon-nucleon
and three-nucleon 
potentials~\cite{Wiringa:2000gb,Pieper:2004qw,Wiringa:2013fia,Pastore:2014oda}. 
The latest work, Ref.~\cite{Pastore:2014oda}, uses the newer AV18+IL7 potential. 

Isospin mixing is addressed in the manner of Ref.~\cite{Barker:1966}: the
empirical total (hadronic) widths are used to fix the isospin-mixing of the states
within a particular doublet. That is, for a doublet of spin $J$, 
the physical states (with labels $a$ and $b$) are given by~\cite{Pastore:2014oda} 
\begin{align}
  \Psi_J^a &= \alpha_J \Psi_{J, T=0} + \beta_J \Psi_{J, T=1}  &
  \Psi_J^b &= \beta_J \Psi_{J, T=0} - \alpha_J \Psi_{J, T=1} \,,	
  \label{mix}
\end{align}
where $a$ denotes 
the lower energy state. Note that $\alpha_J$ and $\beta_J$ are real and satisfy 
$\alpha_J^2 + \beta_J^2=1$. The widths of the isospin-pure states
are computed using the QMC approach, permitting the extraction of 
the mixing parameters in Eq.~(\ref{mix}) from the measured widths, yielding, for example~\cite{Pastore:2014oda},
\begin{align}
	\alpha_1 &= 0.21 (3) 
	&&\text{ and }
	& \beta_1 &= 0.98 (1) \ .
	\label{eq:alpha}
\end{align}
The empirical excitation energies, which are unfolded from the experimental 
data using these mixing coefficients, 
agree with the QMC energies of the states of all three mixed doublets, to within 
the expected theoretical error---that is, to within 1\% uncertainty. 

Given this success, this procedure may be applied to the electromagnetic transitions of these 
isospin-mixed states as well, so that the M1 transitions to the ground state are of the form
\begin{eqnarray}
\langle \Psi_{0,0} || M1  ||  \Psi_J^a \rangle 
&=& \alpha_J M1_{J, T=0} + \beta_J M1_{J, T=1}  \,
\label{eq:M1mix:a}
\\
\langle \Psi_{0,0} || M1  ||  \Psi_J^b \rangle 
&=& \beta_J M1_{J, T=0} - \alpha_J M1_{J, T=1}  \,, 
\label{eq:M1mix:b}
\end{eqnarray} 
where $M1_{J, T}$ is the 
reduced matrix element of the M1 operator with the isospin pure $J, T$ states. 
For reference we note that this matrix element is related to the partial width 
$\Gamma_{M 1}$ for the transition via 
\begin{equation}
\Gamma_{M1} = 
\frac{16\pi}{9} \alpha \hbar c
	\left(\frac{\Delta E}{\hbar c}\right)^{3}
	B(M1) \left(\frac{\hbar c}{2M_p~[\mev]}\right)^{2} \,,
\end{equation}
where 
$B(M1) = |\langle \Psi_{J_f} || M1 || \Psi_{J_i}\rangle|^2 /(2J_i+1)$ is in units of $(\mu_N)^{2}$, the squared nuclear magneton. We emphasize that the M1 operator can mediate both isoscalar ($\Delta T=0$) and isovector ($|\Delta T|=1$) transitions. The $J_{|\Delta T|}$ isospin currents are given in \eqref{X:isospin:current}.

Unfortunately, the leading one-body (impulse approximation) 
results compare poorly to experiment.  The inclusion of  
meson-exchange currents in the $M1_{J, T}$ matrix element improves matters
considerably, 
yielding finally $\Gamma_{M1} = 12.0 (3)$ eV for the 17.64 MeV transition, 
to be compared with 
$\Gamma_{M1}^{\rm expt} = 15.0 (1.8)$ eV~\cite{Tilley:2004zz}, and  
$\Gamma_{M1} = 0.50 (2)$ eV for the 18.15 MeV transition, to be compared with
$\Gamma_{M1}^{\rm expt} = 1.9 (4)$ eV~\cite{Tilley:2004zz}. 
Nevertheless, the discrepancies are still significant, and it would seem that something is missing. It is possible that the treatment of wave function mixing is somehow inadequate. Table V of Ref.~\cite{Pastore:2014oda} shows that increasing the value of $\alpha_1$ to 0.31 makes the M1 transition rate of the 18.15~\mev state double, while decreasing the 17.64~\mev transition by only 5\%~\cite{wiringa}. 

The deficiency can be redressed in a distinct way that has not previously been considered in this context. Isospin breaking can appear in the hadronic form of the electromagnetic transition operators themselves~\cite{Gardner:1995uq,Gardner:1995ya} to the end that changes in the relative strength of the isoscalar and isovector transition operators appear as a result of isospin-breaking in the masses of isospin multiplet states, such as the  nonzero neutron-proton mass difference. This is pertinent because electromagnetic transition operators involve both one and two-body contributions.  The nuclear structure calculations of Ref.~\cite{Pastore:2014oda} employ electromagnetic transition operators from chiral effective theory in the isospin limit~\cite{Pastore:2011ip,Piarulli:2012bn}. The empirical magnetic moments of the neutron and proton are employed in the leading one-body terms in these analyses, albeit they are normalized by the average nucleon mass, rather than the proton mass that appears in the definition of the nuclear magneton. Consequently the isospin-breaking effects that shift the relative strength of the isoscalar and isovector transition operators appear in higher-order terms, namely in the relativistic corrections to leading one-body operators, as well as in the two-body operators. These effects are likely numerically important for the dominantly isoscalar electromagnetic transitions because the relativistic one-body corrections and two-body contributions are predominantly isovector in the isospin limit~\cite{Pastore:2014oda,Pastore}, though technically these corrections to a given contribution appear in higher order in the chiral expansion. 

We choose to include these isospin-breaking effects through the use of a spurion formalism~\cite{lee1981particle}. That is, we include isospin-breaking contributions through the introduction of a fictitious particle, the spurion, whose purpose is to allow the inclusion of isospin-breaking effects within an isospin-invariant framework. Since the largest effects should stem from the neutron-proton mass difference, the spurion acts like a new $\Delta T=1$ operator because its size is controlled by $(M_n - M_p)/M_N$, where $M_N$ is the nucleon mass. Since the isoscalar transition operators are extremely small  we include the ``leakage'' of the dominant isovector operators into the isoscalar channel only. This is justified by noting that Ref.~\cite{Pastore:2014oda} used states of pure isospin and included meson exchange currents, to determine the isovector and isoscalar M1 transition strengths to be 
\begin{align}
	M1_{1,T=1} &= 0.767(9)\,\mu_N 
	&&\text{ and }
	&
M1_{1,T=0} &= 0.014(1)\, \mu_N \ ,
\label{eq:matrixelements}
\end{align}
where the numerical dominance of the isovector M1 transition strength arises from 
that of the empirical isovector anomalous magnetic moment and the charged-pion,  meson-exchange 
contribution, which is isovector. 

Characterizing the strength of the $\Delta T=1$ spurion by 
$\kappa$, the matrix elements of \eqsref{M1mix:a}{M1mix:b} are thus amended by the addition of 
\begin{eqnarray} 
\delta \langle \Psi_{0,0} || M1  ||  \Psi_1^a \rangle 
&= & \alpha_1 \kappa  M1_{1, T=1}  
\\
\delta \langle \Psi_{0,0} || M1  ||  \Psi_1^b \rangle 
&=& \beta_1 \kappa M1_{1, T=1}  \, . 
\label{M1mixdelta}
\end{eqnarray} 
The size of $\kappa$ is controlled by non-perturbative effects.  To illustrate 
its role, we assume that it can be determined 
by demanding that the resulting 
M1 transition rate of the 17.64~\mev decay reproduces its experimental value. 
The final M1 transition matrix elements thus read 
\begin{eqnarray}
\langle \Psi_{0,0} || M1  ||  \Psi_1^a \rangle 
&=& \alpha_1 M1_{1, T=0} + \beta_1 M1_{1, T=1} + \alpha_1 \kappa
M1_{1, T=1}  \,,  
\\
\langle \Psi_{0,0} || M1  ||  \Psi_1^b \rangle 
&=& \beta_1 M1_{1, T=0} - \alpha_1 M1_{1, T=1}  + \beta_1 \kappa 
M1_{1, T=1} \,. 
\label{M1mixplus}
\end{eqnarray} 
The needed shift in the M1 partial width of the 17.64~\mev transition is  $3.0\pm 2.1 \,{\rm eV}$. Employing the matrix elements of Ref.~\cite{Pastore:2014oda}, 
we find the central value of $\kappa=0.549$, to yield 
$\langle \Psi_{0,0} || M1  ||  \Psi_1^b \rangle = 0.265\,\mu_N$ and 
a M1 partial width of 
$1.62\,{\rm eV}$, which is within $1\sigma$ of the experimental result. 

With the above discussion of both isospin mixing and isospin breaking in hand, we now turn to their implications for an 
M1 transition mediated by an $X$ boson with vector couplings 
$\varepsilon_n e$ and $\varepsilon_p e$ to the neutron and proton, 
respectively.  The M1 transition mediated by $X$ is 
\begin{equation}
\langle \Psi_{0,0} || M1_X  ||  \Psi_1^b \rangle 
= 
(\varepsilon_n + \varepsilon_p) 
\beta_1 M1_{1, T=0}  + 
(\varepsilon_p - \varepsilon_n) 
(- \alpha_1 M1_{1, T=1}  + \beta_1 \kappa M1_{1, T=1}) \, , 
\end{equation} 
where the neutron and proton $X$ couplings appear because the $^8$Be system contains
equal numbers of neutrons and protons. 
The resulting ratio of partial widths is, then,
\begin{align}
\frac{\Gamma_X}{\Gamma_\gamma} =
\frac{| (\varepsilon_p + \varepsilon_n) \beta_1 M1_{1, T=0}  + (\varepsilon_p - \varepsilon_n) (-\alpha_1 M1_{1, T=1}  + \beta_1 \kappa M1_{1, T=1} ) |^2}
{|  \beta_1 M1_{1, T=0}  - \alpha_1 M1_{1, T=1}  + \beta_1 \kappa M1_{1, T=1}|^2} 
\frac{|\mathbf{k}_X|^3}{|\mathbf{k}_\gamma|^3} \ .
\label{eq:widthratioisospin}
\end{align}
In the limit of no isospin mixing ($\alpha_1 = 0$, $\beta_1 = 1$) and no isospin breaking ($\kappa = 0$), \eqref{widthratioisospin} reproduces \eqref{IPCC:for:X}.  However, substituting the isospin mixing parameters of \eqref{alpha} and the M1 transition strengths of \eqref{matrixelements}, we find 
\begin{align}
\frac{\Gamma_X}{\Gamma_\gamma} 
&=
| -0.09 \, (\varepsilon_p + \varepsilon_n)   + 1.09 \, (\varepsilon_p - \varepsilon_n) |^2
\frac{|\mathbf{k}_X|^3}{|\mathbf{k}_\gamma|^3} 
& \kappa &= 0 
\label{eq:widthratioisospinnumerical:1}
\\
\frac{\Gamma_X}{\Gamma_\gamma} 
&=
|\phantom{+}\; 0.05 \, (\varepsilon_p + \varepsilon_n)   + 0.95 \, (\varepsilon_p - \varepsilon_n) |^2
\frac{|\mathbf{k}_X|^3}{|\mathbf{k}_\gamma|^3} 
& \kappa &= 0.549 \ .
\label{eq:widthratioisospinnumerical}
\end{align}
The isoscalar contribution is only a small fraction of the isovector one, and so, in general, large modifications from isospin violation are possible. 

\begin{figure}[tb]
\includegraphics[width=\linewidth]{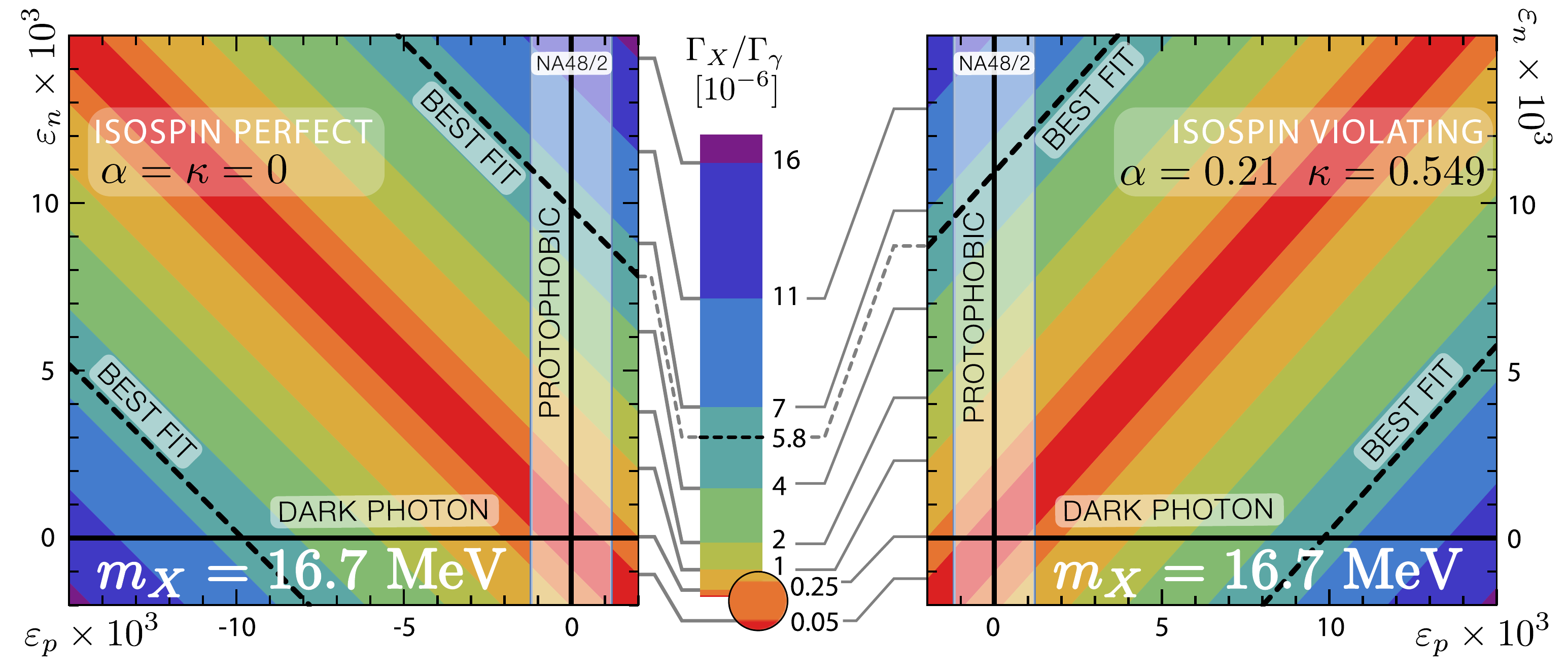} 
\vspace*{-0.3in}
\caption{The ratio $\Gamma_X / \Gamma_\gamma$ in the case of perfect isospin ($\alpha_1 = \kappa = 0$) (left) and isospin violation ($\alpha_1 = 0.21$, $\kappa = 0.549$) (right) in the $(\varepsilon_p, \varepsilon_n)$ plane for $m_X = 16.7~\mev$. The effects of isospin violation may be significant in general, but for the viable protophobic regions of parameter space consistent with NA48/2 constraints (shaded), their effects are small. The best fit value of $\Gamma_X / \Gamma_\gamma = 5.8 \times 10^{-6}$ is highlighted. The dark photon scenario corresponds to $\varepsilon_n=0$.
\label{fig:isospin}}
\end{figure}

In \figref{isospin}, we plot the ratio $\Gamma_X / \Gamma_\gamma$ in the $(\varepsilon_p, \varepsilon_n)$ plane.  In the case of perfect isospin, the transition is isoscalar and the ratio depends on $\varepsilon_p + \varepsilon_n$, but in the case of isospin violation, the isovector transition dominates, and the ratio depends effectively on $\varepsilon_p - \varepsilon_n$.  The effects of including isospin violation are, therefore, generally significant.  Interestingly, however, in the protophobic limit with $\varepsilon_p = 0$, isospin violation only modifies $\Gamma_X / \Gamma_\gamma$ by a factor of about 20\%. However, for larger values of $|\varepsilon_p|$, for example, $|\varepsilon_p| \sim |\varepsilon_n| / 2$, isospin-breaking effects can be significant, leading to factors of 10 changes in the branching ratios, or factors of 3 modifications to the best fit couplings. Such large excursions from protophobia are excluded by the NA48/2 limits for the best fit values of the couplings corresponding to $m_X = 16.7~\mev$, but may be possible for larger values of $m_X$ within its allowed range, as we discuss below.

\section{Signal Requirements for Gauge Boson Couplings}
\label{sec:experimental:effects}

In this section, we discuss what a gauge boson's couplings must be to explain the $^8$Be signal.  We begin with the leptonic couplings, where the requirements are straightforward to determine.  To produce the IPC signal, the $X$ boson must decay to $e^+e^-$.  The Atomki pair spectrometer has a distance of $\mathcal O(\text{few})~\text{cm}$ between the target, where the $^8$Be excited state is formed, and the detectors that observe the charged particles~\cite{Gulyas:2015mia}. The $X$ boson decay width to electrons is
\begin{align}
\Gamma (X \to e^+ e^-) &= \varepsilon_e^2 \alpha \frac{m_X^2 + 2 m_e^2}{3 m_X} 
\sqrt{1 - 4 m_e^2 / m_X^2 } \ ,
\label{eq:decay:width:to:electrons}
\end{align}
with similar formulae for other fermion final states~\cite{Pospelov:2008zw}. Requiring that the new boson propagates no more than 1~cm from its production point implies a lower bound 
\begin{align}
\frac{|\varepsilon_e|}{\sqrt{\text{Br}(X\to e^+e^-)}}
\gtrsim 
{1.3 \times 10^{-5}} \ .	
\label{eq:varpeps_e:within:1cm}
\end{align}
If the $X$ boson couples only to the charged SM fermions required to explain the $^8$Be anomaly, one has $\text{Br}(X\to e^+e^-) = 1$. Note, however, that if $\varepsilon_\nu \neq 0$ or if there exist light hidden-sector states with $X$ charge, then there are generically other decay channels for $X$.

The required quark couplings are determined by the signal event rate, that is, the best fit $\Gamma_X/\Gamma_\gamma$. In the Atomki experimental paper, the best fit branching fraction is that given in \eqref{Krasznahorkay:result}.  Combining this result with the isospin-conserving expression for the branching ratio of \eqref{IPCC:for:X}, we find 
\begin{align}
	&|\varepsilon_p + \varepsilon_n| \approx 
	\frac{1.0 \times 10^{-2}}{\sqrt{\text{Br}(X\to e^+e^-)}}
	&&\text{ or } &
	| \varepsilon_u + \varepsilon_d | 
	&\approx 
	\frac{3.3 \times 10^{-3}}{\sqrt{\text{Br}(X\to e^+e^-)}}
		\ ,
\label{eq:quarkcharges}
\end{align}
where we have taken $m_X = 16.7$~\mev. These results, shifted slightly to $m_X = 17~\mev$, were presented previously in Ref.~\cite{Feng:2016jff}.

Given the discussion above, however, several refinements are in order.  First, one can include the isospin-violating effects discussed in \secref{nuclear:physics:effects}.  These modify the branching ratio expression from \eqref{IPCC:for:X} to \eqref{widthratioisospinnumerical}, with the effects shown in \figref{isospin}.  

\begin{figure}[tb] 
\includegraphics[width=\linewidth]{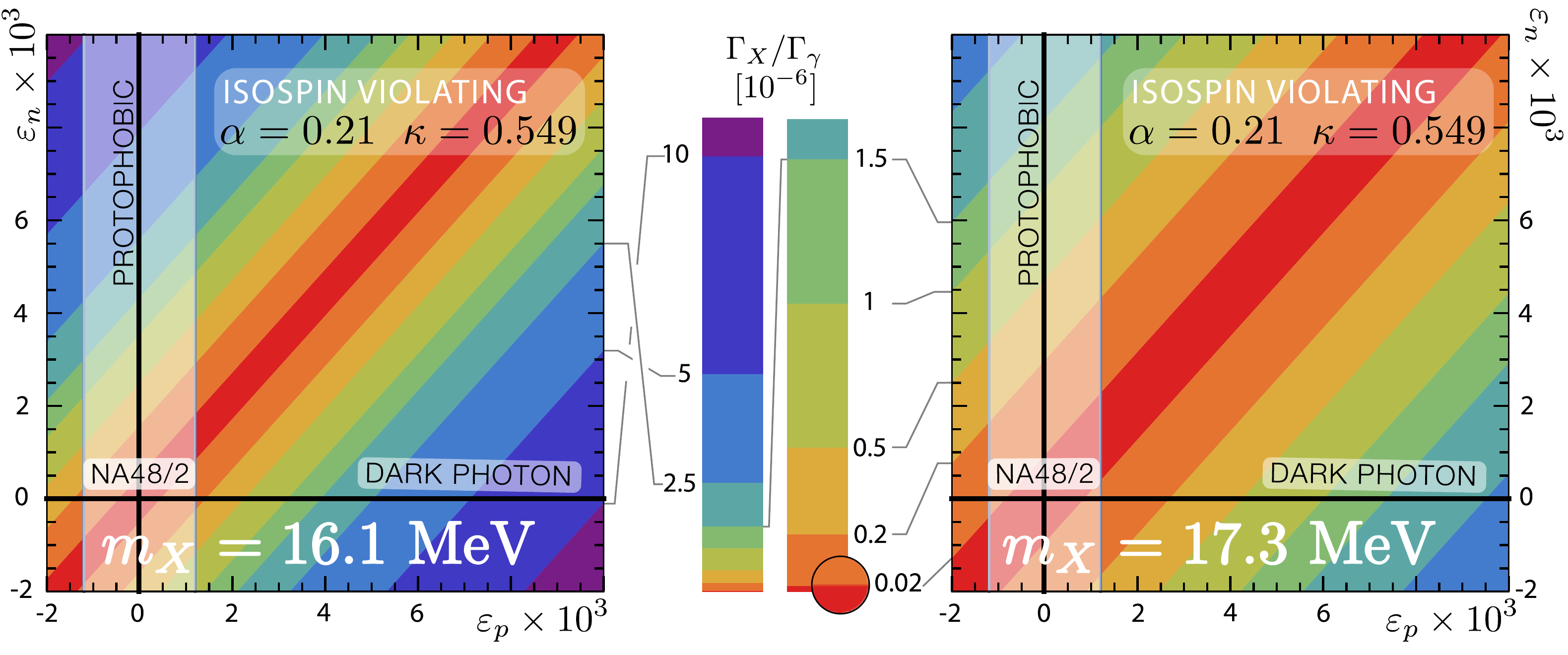}
\vspace*{-0.1in}
\caption{Contours of $\Gamma_X/\Gamma_\gamma$ in the $(\varepsilon_p, \varepsilon_n)$ plane for the parameterization of isospin violation in \eqref{widthratioisospinnumerical}. Also shown are the dark photon axis $(\varepsilon_n = 0)$ and the protophobic region with $|\varepsilon_p| \le 1.2 \times 10^{-3}$ allowed by NA48/2 constraints on $\pi^0 \to X \gamma$. The $m_X$ values are fixed to $m_X =16.1~\mev$ (left) and 17.3~\mev (right), corresponding to the $\pm 1\sigma$ (statistical) range of $m_X$. 
\label{fig:favoredregions:minmax}}  
\end{figure}

Second, as discussed above, the presence of significant isospin mixing strongly suggests that the absence of anomalous IPC decays in the $\Bestarprime$ state originates from kinematic suppression, rather than from isospin symmetry or some other dynamical effect.  This, then, argues for masses in the upper region of the allowed range of \eqref{Krasznahorkay:result}.  Larger masses imply larger phase-space suppression, and these may significantly shift the contours of $\Gamma_X/\Gamma_\gamma$ in the $(\varepsilon_p, \varepsilon_n)$ plane, as can be seen by comparing the $\pm 1\sigma$ values of $m_X$ in  \figref{favoredregions:minmax}.  

Last, and most importantly, to determine the favored couplings, one must know how the best fit $\Gamma_X/\Gamma_\gamma$ depends on $m_X$.  In the original experimental paper, the best fit branching ratio 
$\Gamma_X/\Gamma_\gamma = 5.8 \times 10^{-6}$ was presented without uncertainties and only for the best fit mass of 16.7 MeV.  In a subsequent analysis, however, the experimental collaboration explored the implications of other masses~\cite{privatecommunication}. In preliminary results from this analysis, the M1 and E1 background normalizations were fit to the angular spectrum in the range $40^\circ \le \theta \le 120^\circ$, and confidence regions in the $(m_X, \Gamma_X/\Gamma_\gamma)$ plane were determined with only statistical uncertainties included.  For masses larger than 16.7~\mev, the best fit branching ratio was found to be significantly smaller.  For example, for $m_X = 17.3~\mev$ (17.6~\mev), the best fit was for $\Gamma_X/\Gamma_\gamma \approx 2.3 \times 10^{-6}$ ($0.5 \times 10^{-6}$)~\cite{privatecommunication}.  For such large masses, the best fit with fixed backgrounds is not very good, and the implications for nucleon-level couplings are partially offset by the reduced phase space factor $| \mathbf{k}_X |^3 / | \mathbf{k}_\gamma |^3$. In a full analysis, one should also include systematic errors which are clearly a significant source of uncertainty in the $m_X$ determination, and also let the background levels float in the fit. We expect that including these effects will significantly improve the fit for larger masses and favor even smaller couplings. Specifically, since the anomalous events at angles between 120$^\circ$ and 135$^\circ$ cannot come from signal when the $X$ mass is heavier, larger M1 and E1 backgrounds will improve the fit and thus require smaller signal to achieve the best fit to the angular spectrum.

Clearly a complete understanding of the experimental uncertainties requires a detailed analysis that incorporates an accurate estimate of nuclear isospin violation, simulation of the experiment, systematic uncertainties, varying backgrounds, and the null $\Bestarprime$ result.  Such an analysis is beyond the scope of this study. As a rough estimate of the hadronic couplings required to explain the $^8$Be signal, we take
\begin{align}
	|\varepsilon_n|  &= (2-10) \times 10^{-3} 
	\\ | \varepsilon_p | &\le 1.2 \times 10^{-3} \ ,
\end{align}
where the upper part of the $\varepsilon_n$ range includes the coupling for the best fit branching ratio for $m_X = 16.7~\mev$, and the lower part presumably includes the best fit value for the larger $m_X$ that simultaneously explain the $\Bestar$ signal and the $\Bestarprime$ null results. The proton coupling constraint follows from the NA48/2 constraints to be discussed in \secref{NA48}.  In presenting our models in \secsref{B}{B-L}, we  leave the dependence on $\varepsilon_n$ explicit so that the impact of various values of $\varepsilon_n$ can be easily evaluated.  Note that the lower values of $\Gamma_X/\Gamma_\gamma$ are still too large to accommodate a dark photon explanation.

\section{Constraints From Other Experiments}
\label{sec:constraints}

We now discuss the constraints on the gauge boson's couplings from all other experiments, considering quark, electron, and neutrino couplings in turn, with a summary of all constraints at the end of the section.  Many of these constraints were previously listed in Ref.~\cite{Feng:2016jff}. We discuss them here in more detail, update some---particularly the neutrino constraints---to include new cases and revised estimates from other works, and include other constraints.

\subsection{Quark Coupling Constraints} 
\label{sec:eps:q:bounds}

The production of the $X$ boson in \bestar decays is completely governed by its couplings to hadronic matter. The most stringent bound on these couplings in the $m_X\approx 17$~\mev mass range is the decay of neutral pions into $X\gamma$. For completeness, we also list the leading subdominant constraints on $\varepsilon_q$, for $q=u,d$.

\subsubsection{Neutral pion decay, $\pi^0 \to X\gamma$}
\label{sec:NA48}

The primary constraint on new gauge boson couplings to quarks comes from the NA48/2 experiment, which performs a search for rare pion decays $\pi^0 \to \gamma (X\to e^+e^-)$~\cite{Raggi:2015noa}. 
The bound scales like the anomaly trace factor $N_\pi \equiv (\varepsilon_u q_u - \varepsilon_d q_d)^2$. Translating the dark photon bound $N_{\pi} < \varepsilon_{\text{max}}^2 / 9$ to limits on the new gauge boson couplings gives
\begin{align}
|2 \varepsilon_u + \varepsilon_d |
\;= \;
|\varepsilon_p |
\;\lesssim\;
\frac{ (0.8 - 1.2) \times 10^{-3}}{\sqrt{\text{Br}(X\to e^+e^-)}} \ ,
\label{eq:pion}
\end{align}
where the range comes from the rapid fluctuations in the NA48/2 limit for masses near 17~\mev.  
In Ref.~\cite{Feng:2016jff}, we observed that the left-hand side becomes small when the $X$ boson is protophobic---that is, when its couplings to protons are suppressed relative to neutrons.

\subsubsection{Neutron--lead scattering}

A subdominant bound is set from measurements of neutron-nucleus scattering.  The Yukawa potential acting on the neutron is $V(r) = -(\varepsilon_n e)^2 A e^{-m_X r} / (4 \pi r)$, where $A$ is the atomic mass number. Observations of the angular dependence of neutron--lead scattering constrain new, weakly-coupled forces~\cite{Barbieri:1975xy}, leading to the constraint
\begin{align}
	\frac{(\varepsilon_n e)^2}{4\pi} < 3.4 \times 10^{-11} \left(\frac{m_X}{\mev}\right)^4 \ .
\end{align}

\subsubsection{Proton fixed target experiments}

The $\nu$-Cal~I experiment at the U70 accelerator at IHEP sets bounds from $X$-bremsstrahlung off the initial proton beam~\cite{Blumlein:2013cua} and $\pi^0 \to X\gamma$ decays~\cite{Blumlein:2011mv}. Both of these processes are suppressed in the protophobic scenario so that these bounds are automatically satisfied when \eqref{pion} is satisfied.

\subsubsection{Charged kaon and $\phi$ decays}
There are also bounds on second generation couplings.  The NA48/2 experiment places limits on $K^+ \to \pi^+ (X \to e^+ e^-)$~\cite{Batley:2015lha}.  For $m_X \approx 17~\mev$, the bound on $\varepsilon_n$ is much weaker than the one from $\pi^0$ decays in \eqref{pion}~\cite{Pospelov:2008zw,Davoudiasl:2014kua}.   The KLOE-2 experiment searches for $\phi \to \eta (X \to e^+ e^-)$ and restricts~\cite{Babusci:2012cr}
\begin{align}
	|\varepsilon_s| \lesssim  \frac{1.0 \times 10^{-2}}{\sqrt{\text{Br}(X\to e^+e^-)}} \ .
\end{align}
In principle $\varepsilon_s$ is independent and need not be related to the \bestar coupling. However, in the limit of minimal flavor violation, one assumes $\varepsilon_d = \varepsilon_s$.

\subsubsection{Other meson and baryon decays}

The WASA-at-COSY experiment also sets limits on quark couplings based on neutral pion decays. It is both weaker than the NA48/2 bound and only applicable for masses heavier than  $20~\mev$~\cite{Adlarson:2013eza}.  The HADES experiment searches for dark photons in $\pi^0$, $\eta$, and $\Delta$ decays and restricts the kinetic mixing parameter to $\varepsilon \lesssim 3 \times 10^{-3}$ but only for masses heavier than 20~\mev~\cite{Agakishiev:2013fwl}. HADES is able to set bounds on gauge bosons around 17~\mev in the $\pi^0 \to X X \to e^+ e^- e^+ e^-$ decay channel. This, however, is suppressed by $\varepsilon_n^4$ and is thus insensitive to $| \varepsilon_n | \lesssim 10^{-2}$. Similar considerations suppress $X$ contributions to other decays, such as $\pi^+ \to \mu^+ \nu_{\mu} e^+ e^-$, to undetectable levels.

\subsection{Electron Coupling Constraints}
\label{sec:eps:e:bounds}

The $X$ boson is required to couple to electrons to contribute to IPC events. In \eqref{varpeps_e:within:1cm} we gave a lower limit on $\varepsilon_e$ in order for $X$ to decay within 1~cm of its production in the Atomki apparatus. In this section we review other bounds on this coupling.

\subsubsection{Beam dump experiments}

Electron beam dump experiments, such SLAC E141~\cite{Riordan:1987aw,Bjorken:2009mm}, search for dark photons bremsstrahlung from electrons that scatter off target nuclei. For $m_X = 17$~\mev, these experiments restrict $|\varepsilon_e|$ to live in one of two regimes: either it is small enough to avoid production, or large enough that the $X$ decay products are caught in the dump~\cite{Essig:2013lka}, leading to
\begin{align}
	|\varepsilon_e| &< 10^{-8}
	&\text{or}
	&&
	\frac{|\varepsilon_e|}{\sqrt{\text{Br}(X\to e^+e^-)}}
	 \gtrsim 2\times {10^{-4}} \ .
\end{align}
The region  $| \varepsilon_e | < 10^{-8}$ is excluded since the new boson would not decay inside the Atomki apparatus. This leads to the conclusion that $X$ must decay inside the beam dump.  Less stringent bounds come from Orsay~\cite{Davier:1989wz} and the SLAC E137~\cite{Bjorken:1988as} experiment. The E774 experiment at Fermilab is only sensitive to $m_X < 10~\mev$~\cite{Bross:1989mp}.

\subsubsection{Magnetic moment of the electron}

The upper limit on $|\varepsilon_e|$ can be mapped from dark photon searches that depend only on leptonic couplings. The strongest bound for $m_X = 17$~\mev is set by the anomalous magnetic moment of the electron, $(g-2)_e$, which constrains the coupling of the new boson to be~\cite{Davoudiasl:2014kua}
\begin{align}
|\varepsilon_e| < 1.4 \times 10^{-3}  \ .
\label{eq:magnetic:moment:electron}
\end{align}

\subsubsection{Electron--positron annihilation into $X$ and a photon, $e^+ e^- \to X \gamma$}

A similar bound arises from the KLOE-2 experiment, which looks for $e^+ e^- \to X \gamma$ followed by $X \to e^+ e^-$, and finds $| \varepsilon_e | {\sqrt{\text{Br}(X\to e^+e^-)}} < 2 \times 10^{-3}$~\cite{Anastasi:2015qla}.  An analogous search at BaBar  is limited to $m_X > 20~\mev$~\cite{Lees:2014xha}.

\subsubsection{Proton fixed target experiments}

The CHARM experiment at CERN also bounds $X$ couplings through its searches for $\eta, \eta' \to \gamma (X \to e^+e^-)$~\cite{Gninenko:2012eq}.  The production of the $X$ boson in the CHARM experiment is governed by its hadronic couplings. The couplings required by the anomalous IPC events, \eqref{quarkcharges}, are large enough that the $X$ boson would necessarily be produced in CHARM. Given the lower bound from decay in the Atomki spectrometer, \eqref{varpeps_e:within:1cm}, the only way to avoid the CHARM constraint for $m_X = 17~\mev$ is if the decay length is short enough that the $X$ decay products do not reach the CHARM detector. The dark photon limit on $\varepsilon$ applies to $\varepsilon_e$ and yields 
\begin{align}
\frac{| \varepsilon_e |}{\sqrt{\text{Br}(X\to e^+e^-)}} > 
{2 \times 10^{-5}}
\ .
\end{align}
This is weaker than the analogous lower bound on $|\varepsilon_e|$ from beam dump experiments. LSND data imposes an even weaker constraint~\cite{Athanassopoulos:1997er,Batell:2009di,Essig:2010gu}.

\subsubsection{Charged kaon and $\phi$ decays}
In charged kaon decay to leptons, the $X$ vector boson 
may be emitted from a charged lepton line. Since the new vector interaction does
not respect the precise gauge invariance of the SM, the interaction of the longitudinal component
of $X$ is not constrained by a corresponding conserved current and thus can be 
significantly enhanced
with energy~\cite{Barger:2011mt,Carlson:2012pc,Laha:2013xua,Karshenboim:2014tka}. 
However, the most severe existing limit comes from the nonobservation of 
an excess in $\Gamma(K \to \mu + {\rm inv.})$ with respect to 
$\Gamma(K\to \mu \nu)$~\cite{Barger:2011mt,Carlson:2012pc,Laha:2013xua}, which is not
pertinent here as we require an appreciable $\Gamma(X\to e^+ e^-)$ 
in order to explain 
the $^8$Be anomaly. 

\subsubsection{$W$ and $Z$ decays}

The $X$ boson can be produced as final state-radiation in $W$ and $Z$ decays into SM fermions. When the $X$ then decays into an electron--positron pair, this gives a contribution to $\Gamma (Z\to  4e)$ that is suppressed by $\mathcal O(\varepsilon_e^2)$. For the electron couplings $\varepsilon_e \alt 10^{-3}$ required here, the impact on the inclusive widths is negligible compared to the order per mille experimental uncertainties on their measurement~\cite{Amsler:2008zzb}.  The specific decay $Z \to 4\ell$ has been measured to lie within $10\%$ of the SM expectation by ATLAS and CMS~\cite{CMS:2012bw,Aad:2014wra} and is consistent with the couplings of interest here.

A more severe constraint arises, however, from the experimental value of the $W$ width, 
because the enhancement mentioned in leptonic $K$ decay appears in 
$W\to \mu \nu X$ as well~\cite{Karshenboim:2014tka}. Limiting the contribution of 
$W \to \ell \nu X$ to twice the error in the $W$ width, after 
Ref.~\cite{Karshenboim:2014tka}, yields
\begin{equation}
|\varepsilon_{\ell}| < 4.2 \times 10^{-3} \left( \frac{m_X}{\hbox{10 MeV}} \right) \,,
\end{equation}
to leading order in $m_X/m_W$ and $m_\ell/M_W$, 
where we have assumed lepton universality and that $\ell \in e, \mu, \tau$ can contribute
to the $W$ width. The resulting constraint on $\varepsilon_e$ is weaker than that
from the magnetic moment of the electron.

\subsection{Neutrino Coupling Constraints} 
\label{sec:eps:nu:bounds}

The interaction of a light gauge boson with neutrinos is constrained in multiple ways, depending on the SM currents to which the boson couples; see Refs.~\cite{Boehm:2003hm,Boehm:2004uq,Laha:2013xua, Jeong:2015bbi}.  The neutrino coupling is relevant for the $^8$Be anomaly because SU(2)$_L$ gauge invariance relates the electron and neutrino couplings. Because neutrinos are lighter than electrons, this generically opens additional $X$ decay channels and reduces $\text{Br}(X\to e^+e^-)$. This, in turn, reduces the lower bound on $\varepsilon_e$ in \eqref{varpeps_e:within:1cm} and alleviates many of the experimental constraints above at the cost of introducing new constraints from $X$--neutrino interactions.

\subsubsection{Neutrino--electron scattering}

Neutrino--electron scattering stringently constrains the $X$ boson's leptonic couplings~\cite{Bilmis:2015lja,Khan:2016uon}.  In the mass range $m_X \approx 17~\mev$, the most stringent constraints are from the TEXONO experiment, where $\bar{\nu}_e$ reactor neutrinos with average energy $\langle E_{\nu} \rangle = 1-2~\mev$ travel 28 meters and scatter off electrons. The resulting electron recoil spectrum is measured.  The path length is short, so the neutrinos remain in nearly pure $\nu_e$ flavor eigenstates.  In the SM, $\bar{\nu}_e e \to \bar{\nu}_e e$ scattering is mediated by both $s$- and $t$-channel diagrams. A new neutral gauge boson that couples to both neutrinos and electrons induces an additional $t$-channel contribution. 

Because constraints from $\bar{\nu}_e e$ scattering are sensitive to the interference of SM and new physics, they depend on the signs of the new gauge couplings, unlike all of the other constraints discussed above.  The importance of the interference term has been highlighted in Ref.~\cite{Bilmis:2015lja} in the context of a $B-L$ gauge boson model.  In that model, the neutrino and electron couplings have the same sign, and the interference was found to be always constructive. 

Assuming that the experimental bound is determined by the total cross section and not the shape of the recoil spectrum, one may use the results of Ref.~\cite{Bilmis:2015lja} to determine the bounds in our more general case, where the couplings can be of opposite sign and the interference may be either constructive or destructive.  Define the quantity $g \equiv |\varepsilon_e\varepsilon_\nu|^{1/2}$.  Let $\Delta \sigma$ be the maximal allowed deviation from the SM cross section and $g_\pm$ ($g_0$) be the values of $g$ that realize $\Delta \sigma$ in the case of constructive/destructive (negligible) interference,
\begin{align}
	\Delta \sigma &= \phantom{+} g_0^4 \sigma_X
	\\
	\Delta \sigma &= \phantom{+} g_+^2 \sigma_\text{int} + g_+^4 \sigma_X
	\\
	\Delta \sigma &= -g_-^2 \sigma_\text{int} + g_-^4 \sigma_X \ ,
\end{align}
where $g^4 \sigma_X$ is the purely $X$-mediated contribution to the cross section and $g^2\sigma_\text{int}$ is the absolute value of the interference term.  Solving these equations for the $g$'s yields the simple relation
\begin{equation}
g_- g_+ = g_0^2 \ .
\end{equation}
The authors of Ref.~\cite{Bilmis:2015lja} found that for $m_X = 17~\mev$, the maximal allowed $B-L$ gauge boson coupling, $g_{B-L}$, is $2\times 10^{-5}$ and $4 \times 10^{-5}$ in the cases of constructive interference and no interference, respectively.  From this, including the factor of $e$ difference between the definitions of $g_{B-L}$ and our $\varepsilon$'s, we find 
\begin{align}
\sqrt{| \varepsilon_e \varepsilon_{\nu} |} & < 7 \times 10^{-5}  
&
\text{for}\quad
\varepsilon_e \varepsilon_{\nu} & > 0 
\quad \text{(constructive interference)} 
\\
\sqrt{| \varepsilon_e \varepsilon_{\nu} |} &< 3 \times 10^{-4} 
&
\text{for}\quad
\varepsilon_e \varepsilon_{\nu} & < 0 
\quad \text{(destructive interference)} \ .	
\end{align}
The relative sign of the couplings thus has a significant effect.  For a fixed value of $\varepsilon_e$, the bound on $|\varepsilon_{\nu}|$ is 16 times weaker for the sign that produces destructive interference than for the sign that produces constructive interference.\\

\subsubsection{Neutrino--nucleus scattering}

In addition to its well-known motivations of providing interesting measurements of $\sin\theta_W$ and bounds on heavy $Z'$ boson~\cite{McLaughlin:2015xfa,Dutta:2015vwa}, coherent neutrino--nucleus scattering, may also provide leading constraints on light, weakly-coupled particles~\cite{Williams:2011qb,deNiverville:2015mwa}.  Although $\nu$--$N$ scattering has not yet been observed, it is the target of a number of upcoming experiments that use reactors as sources.  In addition, the process can also be probed using current and next-generation dark matter direct detection experiments by searching for solar neutrino scattering events~\cite{Dent:2016wor}. For a $B-L$ gauge boson, this sensitivity has been estimated in Ref.~\cite{Cerdeno:2016sfi} for SuperCDMS, CDMSlite, and LUX, with the latter providing the most stringent constraint of $g_{B-L} \lesssim 1.5 \times 10^{-4}$. Rescaling this result to the case of a boson with couplings $\varepsilon_\nu e$ and $\varepsilon_{p,n} e$ to nucleons yields
\begin{align}
	\varepsilon_\nu \varepsilon_n \left[(A-Z) + Z\frac{\varepsilon_p}{\varepsilon_n}\right] < \frac{A}{4\pi\alpha} \left(1.5 \times 10^{-4}\right)^2 \ ,
\end{align}
where we approximate the LUX detector volume to be composed of a single xenon isotope. Since the NA48/2 bounds on $\pi^0\to X\gamma$ imply the protophobic limit where $\varepsilon_p \ll \varepsilon_n$, the second term on the left-hand side may be ignored. Taking $A = 131$ and $Z=54$ then yields $| \varepsilon_{\nu} \varepsilon_n |^{1/2} < 6 \times 10^{-4}$ or
\begin{align}
	\varepsilon_\nu < 2\times 10^{-4} \left(\frac{0.002}{\varepsilon_n}\right) \ .
\end{align}
This bound is weaker than the $\nu$--$e$ scattering bound with constructive interference and comparable to the $\nu$--$e$ bound with destructive interference.  As the $\nu$--$N$ bounds are estimated sensitivities, we use the $\nu$--$e$ bounds in the discussion below.

\subsection{Summary of Constraints}

\begin{figure}[t]
\includegraphics[height=.45\linewidth]{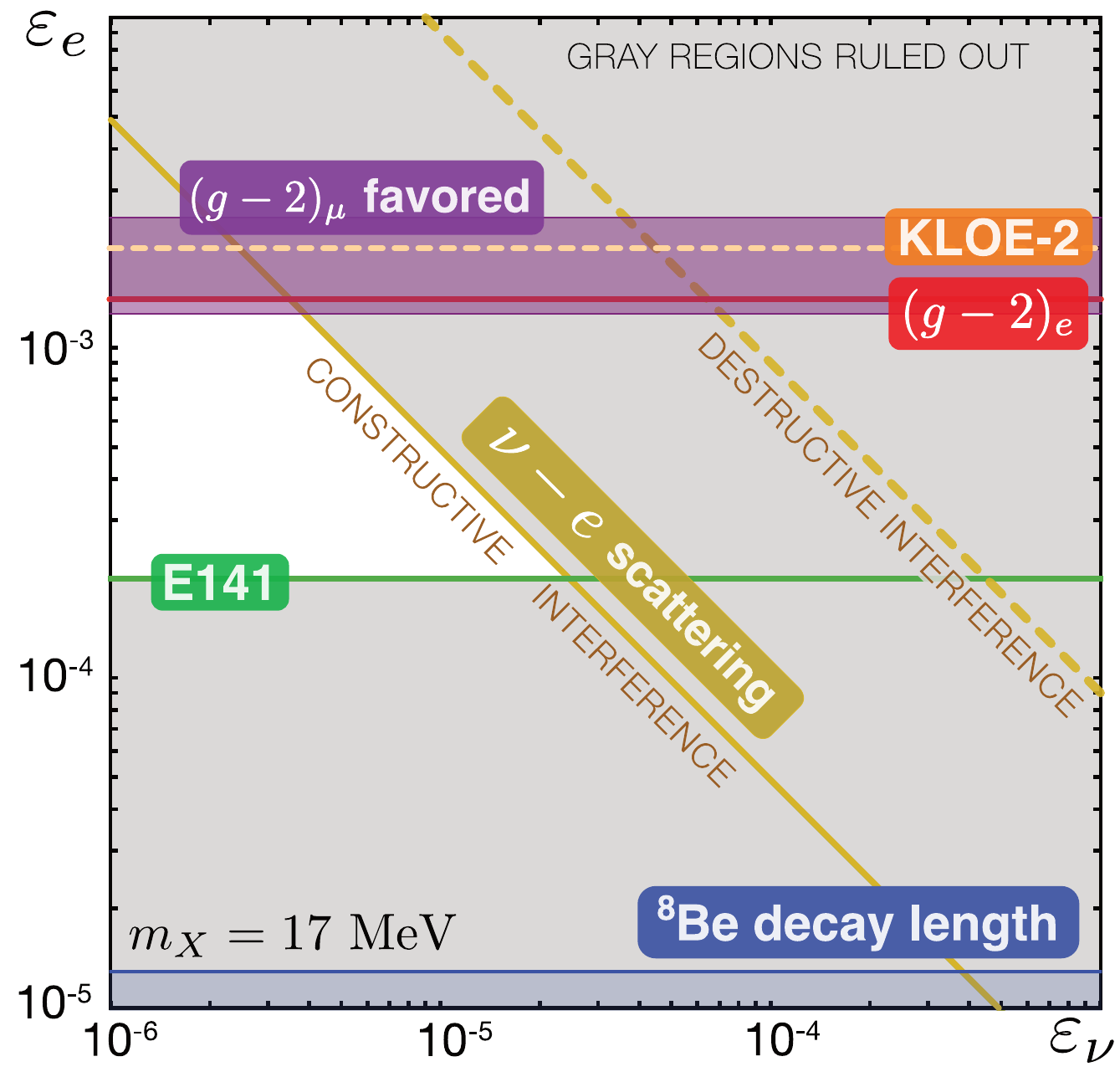} 
\vspace*{-0.1in}
\caption{Summary of constraints and target regions for the leptonic couplings of a hypothetical $X$ gauge boson with $m_X \approx 17$~\mev. Updated from Ref.~\cite{Feng:2016jff}.
\label{fig:constraint:summary}} 
\end{figure}

Combining the required ranges of the couplings to explain the $^8$Be signal from \secref{experimental:effects} with the strongest bounds from other experiments derived above, we now have the acceptable ranges of couplings for a viable protophobic gauge boson to explain the $^8$Be signal.  Assuming $\text{Br}(X \to e^+ e^-) = 1$, the requirements are
\begin{eqnarray}
| \varepsilon_n | &\;=\;& (2-10) \times 10^{-3} 
\\
| \varepsilon_p | &\;\alt \; & 1.2 \times 10^{-3} 
\\
| \varepsilon_e | &\;=\;& (0.2 - 1.4) \times 10^{-3} 
\label{eq:lepton:allowed:couplings1}
\\
\sqrt{| \varepsilon_e \varepsilon_{\nu} |} &\;\alt \;& 3 \times 10^{-4}	\ .
	\label{eq:lepton:allowed:couplings}
\end{eqnarray}
The nucleon couplings are fixed to reproduce the $^8$Be signal rate while avoiding the $\pi^0\to X\gamma$ decays, and the quark couplings are related by $\varepsilon_u + 2 \varepsilon_d = \varepsilon_n$ and $2 \varepsilon_u + \varepsilon_d = \varepsilon_p$. The electron coupling is bounded from above by $(g-2)_e$ and KLOE-2 and from below by beam dump searches, and the neutrino coupling is bounded by $\nu$--$e$ scattering.  The allowed lepton coupling regions are shown in \figref{constraint:summary}. 

\section{U(1)$_B$ Model for the Protophobic Gauge Boson}
\label{sec:modelB}

In this section, we present anomaly-free extensions of the SM where the protophobic gauge boson is a light U(1)$_{B}$ gauge boson that kinetically mixes with the photon.  These models have significant virtues, which we identify in \secref{B}. One immediate advantage is that it does not differentiate between left- and right-handed SM fermions, and so naturally has non-chiral couplings.  Depending on the best fit couplings discussed in \secref{experimental:effects}, the resulting models may be extremely simple, requiring only the addition of extra particles to cancel the anomalies, as discussed in \secref{anomalies}.

\subsection{U(1)$_B$ Gauge Boson with Kinetic Mixing}
\label{sec:B}

The promotion of U(1)$_{B}$ baryon number from a global to a local symmetry has recently attracted attention~\cite{FileviezPerez:2010gw,Dulaney:2010dj,FileviezPerez:2011dg,FileviezPerez:2011pt,Duerr:2013dza,Arnold:2013qja,Perez:2014qfa,Duerr:2014wra,Farzan:2016wym}. Gauged U(1)$_B$ is not anomaly-free, but these studies have constructed a number of models in which the gauge anomalies are cancelled with rather minimal new matter content.  

Here we assume that the U(1)$_{B}$ symmetry is broken through a Higgs mechanism, as discussed below, generating a mass for the $B$ gauge boson.  As with all Abelian symmetries, the $B$ gauge boson will generically mix kinetically with the other neutral gauge bosons of the SM. At energies well below the weak scale, this mixing is dominantly with the photon. The resulting Lagrangian is
\begin{eqnarray}
\label{gbasisL}
{\cal L} &=& 
- \frac{1}{4} \tilde{F}_{\mu \nu} \tilde{F}^{\mu \nu}
- \frac{1}{4} \tilde{X}_{\mu \nu} \tilde{X}^{\mu \nu}
+ \frac{\epsilon}{2} \tilde{F}_{\mu \nu} \tilde{X}^{\mu \nu} 
+ \frac{1}{2} m_{\tilde{X}}^2 \tilde{X}_{\mu} \tilde{X}^{\mu}  
+ \sum_f \bar{f} i \slashed{D} f  \ ,
\end{eqnarray}
where $\tilde{F}_{\mu\nu}$ and $\tilde{X}_{\mu\nu}$ are the field strengths of the photon and $B$ gauge boson, the sum runs over all fermions $f$, and the covariant derivative is
\begin{eqnarray}
D_{\mu} &=& \partial_{\mu} + i e Q_f \tilde{A}_{\mu} + i e \epsilon_{B} B_f \tilde{X}_{\mu} \ .
\end{eqnarray}
Here $Q_f$ and $B_f$ are the electric charge and baryon number of fermion $f$, and $\epsilon_{B}$ is the $B$ gauge coupling in units of $e$. The tildes indicate gauge-basis fields and quantities.

In the mass basis, the Lagrangian is 
\begin{eqnarray}
{\cal L} &=&
- \frac{1}{4} F_{\mu \nu} F^{\mu \nu}
- \frac{1}{4} X_{\mu \nu} X^{\mu \nu}
+ \frac{1}{2} m_{X}^2 X_{\mu} X^{\mu} 
+ \sum_f \bar{f} i \slashed{D}_{\mu} f \ ,
\end{eqnarray}
where 
\begin{eqnarray}
m_X \equiv \frac{1}{\sqrt{1-\epsilon^2}} m_{\tilde{X}}
\end{eqnarray}
is the physical $X$ boson mass, and
\begin{align}
\tilde{A}_{\mu} &\equiv A_{\mu} + \frac{\epsilon}{\sqrt{1-\epsilon^2}} X_{\mu} &
\tilde{X}_{\mu} &\equiv \frac{1}{\sqrt{1-\epsilon^2}} X_{\mu} 
\end{align}
define the physical massless photon $A$ and massive gauge boson $X$.  The fermions couple to photons with the usual charge $e Q_f$, but they couple to the $X$ boson with charge $e \varepsilon_f$, where
\begin{eqnarray}
\varepsilon_f = \varepsilon_{B} B_f + \varepsilon Q_f \ ,
\end{eqnarray}
and the script quantities are defined by
\begin{align}
\varepsilon_{B} &= \frac{\epsilon_{B}}{\sqrt{1-\epsilon^2}} 
&
\varepsilon &= \frac{\epsilon}{\sqrt{1-\epsilon^2}} \ .
\end{align}

The $X$ charges for the SM fermions, using 1st generation notation, are
\begin{eqnarray}
\varepsilon_u &=&  \frac{1}{3} \varepsilon_{B} + \frac{2}{3} \varepsilon  \\
\varepsilon_d &=& \frac{1}{3} \varepsilon_{B} - \frac{1}{3} \varepsilon   \\
\varepsilon_\nu &=& 0 \\
\varepsilon_e &=& - \varepsilon \ .
\end{eqnarray}
The $\pi^0$ constraints we have discussed above require $\varepsilon$ and $-\varepsilon_{B}$ to be approximately equal to within 10\% to 50\%.  It is therefore convenient to define $\varepsilon \equiv - \varepsilon_{B} + \delta$, so 
\begin{eqnarray}
\varepsilon_u &=& - \frac{1}{3} \varepsilon_B + \frac{2}{3} \delta  \\
\varepsilon_d &=& \frac{2}{3} \varepsilon_B - \frac{1}{3} \delta  \\
\varepsilon_\nu &=& 0  \\
\varepsilon_e &=& \varepsilon_B - \delta  \ ,
\end{eqnarray}
with corresponding nucleon charges $\varepsilon_n = \varepsilon_B$ and $\varepsilon_p = \delta$.

This model has some nice features. For small $\delta$, the charges are $Q-B$, which satisfies the protophobic condition.  For the same reason, the neutrino's charge is identically zero.  As discussed in \secref{eps:nu:bounds}, the constraints on neutrino charge are among the most stringent, both given $\nu$--$e$ and $\nu$--$N$ constraints, and the $^8$Be signal requirement that $X$ decays not be dominated by the invisible decay $X \to \nu \bar{\nu}$.  The model is highly constrained, and we see that the electron coupling is not suppressed relative to the quark couplings. However, for $\varepsilon_B \approx 0.002$ and $\delta \approx 0.001$, this model provides an extremely simple and minimal explanation of the $^8$Be signal (provided gauge anomalies are cancelled, as discussed below). Note that it predicts values of $\varepsilon_e \approx 0.001$, that is, in the upper part of allowed range of \eqref{lepton:allowed:couplings1}.  Assuming $\varepsilon_{\mu} \approx \varepsilon_e$, such couplings remove~\cite{Pospelov:2008zw} at least part of the longstanding discrepancy in $(g-2)_{\mu}$ between measurements~\cite{Bennett:2006fi} and the SM prediction~\cite{Hagiwara:2006jt}, with important implications for the upcoming Muon $(g-2)$ Experiment at Fermilab~\cite{Grange:2015fou}.  They also imply promising prospects for future searches for the protophobic $X$ boson at low-energy colliders, as discussed in \secref{conclusions}.

We treat the kinetic mixing $\varepsilon$ as a free parameter.  In a more fundamental theory, however, $\varepsilon$ may be related to $\varepsilon_{B}$.  For example, if U(1)$_{B}$ is embedded in non-Abelian gauge group, $\varepsilon$ vanishes above the symmetry-breaking scale, but when the non-Abelian symmetry breaks, it is generated by vacuum polarization diagrams with particles with electric charge and $B$ quantum numbers in the loop. Parametrically, $\varepsilon \sim ({e^2}/{6 \pi^2}) \varepsilon_{B} \sum_f  Q_f B_f \ln r_f$~\cite{Holdom:1985ag}, where the sum is over pairs of particles in the loop, and the $r_f$ are ratios of masses of these particles.  Given $\sim 100$ particles, one would therefore expect $\varepsilon \sim \varepsilon_{B}$ in general, and the particular relation $\varepsilon \approx - \varepsilon_{B}$, which is not renormalization group-invariant, may be viewed as providing information at low-energy scales about the GUT-scale particle spectrum.

\subsection{Anomaly Cancellation and Experimental Implications}
\label{sec:anomalies}

Models with gauged baryon number require additional particle content to cancel anomalies. The simplest experimentally viable extension of the SM with gauged U(1)$_B$ requires adding three vectorlike pairs of color-singlet fields~\cite{Duerr:2013dza,Duerr:2014wra}.\footnote{A model unifying gauged baryon number and color into a non-Abelian SU(4) has been constructed and, after symmetry breaking, yields the same new particle content as the U(1)$_B$ model discussed here~\cite{Fornal:2015boa}.}  These fields and their quantum numbers are listed in \tableref{gaugedB}. The new fields carry baryon charges that satisfy the anomaly cancellation condition $B_2-B_1 = 3$. The $\chi$ field is naturally a dark matter candidate~\cite{Duerr:2013lka,Duerr:2014wra}, and it has to be the lightest of the new fields to avoid stable charged matter. 

\begin{table}[b]
\caption{New particle content of the simplest anomaly-free U(1)$_B$ model.}
 \label{table:gaugedB}
\begin{tabular}[t]{ccccc}
 \toprule 
Field & \ Isospin $I$ \ & \ Hypercharge $Y$ \ & \ $B$ \vspace{0.02in} \\ \hline
$S_B$ & 0 & 0 & 3 \\
         $\Psi_{L}$ & $\frac{1}{2}$ & $-\frac{1}{2}$ & $B_1$  \\ 
         $\Psi_R$ & $\frac{1}{2}$ & $-\frac{1}{2}$ &  $B_2$  \\ 
          $\eta_R$ & $0$ & $-1$ &  $B_1$  \\ 
               $\eta_L$ & $0$ & $-1$ & $B_2$  \\ 
         $\chi_R$ & $0$ & $0$ &  $B_1$  \\ 
          $\chi_L$ & $0$ & $0$ &  $B_2$  \\ 
\toprule
 \end{tabular}
\end{table}

The U(1)$_B$ symmetry is broken by the vacuum expectation value (vev) $\langle S_B \rangle = v_X/\sqrt2$ of a new SM-singlet Higgs field carrying baryon number $B = 3$ to allow for vectorlike mass terms and to make the $\chi$ field the lightest one. The new Yukawa terms in the Lagrangian are
\begin{eqnarray}
 \mathcal{L}_Y &=& 
 - y_1 \overline{\Psi}_L h_{\text{SM}} \eta_R - y_2 \overline{\Psi}_L \tilde{h}_{\text{SM}} \chi_R 
- y_3 \overline{\Psi}_R h_{\text{SM}} \eta_L -y_4 \overline{\Psi}_R \tilde{h}_{\text{SM}} \chi_L  
\nonumber \\
&& - \lambda_\Psi S_B \overline{\Psi}_L \Psi_R - \lambda_\eta S_B \overline{\eta}_R \eta_L 
- \lambda_\chi S_B \overline{\chi}_R \chi_L + {\rm h.c.} 
\label{eq:Yukawas}
\end{eqnarray}
In Refs.~\cite{Duerr:2013dza,Duerr:2014wra} U(1)$_B$ is assumed to be broken at the TeV scale. However, to have a light U(1)$_B$ gauge boson and a gauge coupling consistent with the $^8$Be signal, the vev of the new Higgs boson cannot be so large.
Defining its vacuum expectation value by $\langle S_B \rangle = v_X/\sqrt2$, the mass of the new $X$ gauge boson corresponding to the broken U(1)$_B$ is given by 
\begin{equation}
m_{X} =  3 e |\varepsilon_B| v_X \ ,
\end{equation}
implying
\begin{equation}
v_X \approx 10~\gev \, \frac{0.002}{|\varepsilon_B | }\ .
\label{eq:vX}
\end{equation}
 As a result, the new particles cannot have large vectorlike masses from the $\lambda_i$ couplings in \eqref{Yukawas}, but must rather have large chiral couplings from the $y_i$ terms of \eqref{Yukawas}.

The experimental constraints on the extra matter content of this model come from several sources:
\begin{itemize} 

\item[$\bullet$]  First, the new particles may be produced through Drell-Yan production at the LHC. However, for Yukawa couplings $y_i \sim 3$, close to the perturbative limit, the masses of the new states are $\sim 500~\gev$ and beyond current LHC sensitivity.

\item[$\bullet$]  Second, electroweak precision measurements constrain the properties of the new particles.  The two electroweak doublets give an irreducible contribution to the $S$ parameter of $\Delta S \approx 2/(6\pi) \simeq 0.11$~\cite{Peskin:1991sw}. In the degenerate mass limit, they do not contribute to the $T$ and $U$ parameters. However, the fit to electroweak precision data may be improved with a slight splitting of $\Delta m \sim 50~\gev$, which gives $\Delta T \approx 2/(3\pi\sin^22\theta_W)(\Delta m/m_Z)^2 \approx 0.09$. This combination of $\Delta S$ and $\Delta T$ fits well within the 90\% CL region (see, for example, Fig.~10.6 of Ref.~\cite{Agashe:2014kda}).

\item[$\bullet$]  Third, the new particles may affect the $h_{\text{SM}} \to \gamma\gamma$ decay rate. Since these particles essentially form two additional families of leptons, the rate for Higgs decaying to two photons decreases by $\sim 20\%$ compared to the SM prediction~\cite{Kribs:2007nz}, but this is still within the experimentally-allowed region~\cite{combined}.

\end{itemize}

In summary, a simple model with a U(1)$_B$ gauge boson that kinetically mixes with the photon is a viable candidate for the protophobic gauge boson.  The gauge anomalies must be cancelled by introducing additional particles, and we have discussed the simplest realization of this field content that simultaneously explains the $^8$Be anomaly.

\section{U(1)$_{B-L}$ Model for the Protophobic Gauge Boson}
\label{sec:modelB-L}

In this section, we present another anomaly-free extension of the SM where the protophobic gauge boson is a light U(1)$_{B-L}$ gauge boson that kinetically mixes with the photon.  These models have significant virtues, which we identify in \secref{B-L}.  They also generically have neutrino couplings that are too large, and we explore a mechanism for suppressing the neutrino couplings in \secref{active}.  The resulting models may be extremely simple, requiring only the addition of one generation of vectorlike leptons which is light and may already be probed at the LHC. The implications for colliders and cosmology are discussed in \secref{activeimplications}.

\subsection{U(1)$_{B-L}$ Gauge Boson with Kinetic Mixing}
\label{sec:B-L}

The possibility of gauged U(1)$_{B-L}$ has been studied for many decades~\cite{Mohapatra:1980qe,Buchmuller:1991ce,Buchmuller:1992qc,Heeck:2014zfa}.  The promotion of U(1)$_{B-L}$ from a global to a local symmetry is well-motivated among Abelian symmetries by its appearance in grand unified theories, and the fact that it is anomaly-free once one adds to the SM three right-handed (sterile) neutrinos, which are already strongly motivated by the existence of neutrino masses.  

As in the U(1)$_B$ case, we assume that the $B-L$ symmetry is broken through a Higgs mechanism, generating a mass for the $B-L$ gauge boson, and that it kinetically mixes with the photon.  The resulting $X$-charges for the SM fermions, using 1st generation notation, are
\begin{eqnarray}
\varepsilon_u &=& \frac{1}{3} \varepsilon_{B-L}  + \frac{2}{3} \varepsilon \\
\varepsilon_d &=& \frac{1}{3} \varepsilon_{B-L} - \frac{1}{3} \varepsilon \\
\varepsilon_\nu &=& - \varepsilon_{B-L} \\
\varepsilon_e &=& - \varepsilon_{B-L} - \varepsilon  \ ,
\end{eqnarray}
or, defining $\varepsilon \equiv - \varepsilon_{B-L} + \delta$ as above, 
\begin{eqnarray}
\varepsilon_u &=& - \frac{1}{3} \varepsilon_{B-L} + \frac{2}{3} \delta  \\
\varepsilon_d &=& \frac{2}{3} \varepsilon_{B-L} - \frac{1}{3} \delta  \\
\varepsilon_\nu &=&  - \varepsilon_{B-L}  \\
\varepsilon_e &=& - \delta  \ .
\end{eqnarray}
The corresponding nucleon charges are $\varepsilon_n = \varepsilon_{B-L}$ and $\varepsilon_p = \delta$.  

The charges of the kinetically mixed $B-L$ gauge boson have nice features for explaining the $^8$Be anomaly.  For $\delta \approx 0$, the charges are $Q-(B-L)$, which satisfies the basic requirements of a protophobic solution to the $^8$Be anomaly: namely, the $X$ boson couples to neutrons, but its couplings to both protons and electrons are suppressed.  More quantitatively, by choosing  the two parameters $|\varepsilon_{B-L}| \approx 0.002 - 0.008$ and $|\delta| \alt 0.001$, the up and down quark couplings give the $^8$Be signal and are sufficiently protophobic to satisfy the $\pi^0$ constraints. This is no great achievement: by picking two free parameters, two conditions can be satisfied.  But what is non-trivial is that with this choice, the electron coupling satisfies the upper bound $|\varepsilon_e| \alt 1.4 \times 10^{-3}$, which is required by the completely independent set of experiments that constrain lepton couplings.

Unfortunately, in contrast to the U(1)$_B$ case, the neutrino coupling does not vanish.  In these models, we see that $\varepsilon_{\nu} = -\varepsilon_n$ while the constraints discussed above require the neutrino coupling to be significantly below the neutron coupling.  In the next section, we present a mechanism to neutralize the $X$-charge of SM active neutrinos to satisfy these bounds.

\subsection{Neutrino Neutralization with Vectorlike Leptons}
\label{sec:active}

The $B-L$ gauge boson with kinetic mixing predicts $|\varepsilon_{\nu}| = |\varepsilon_n| \sim 0.002 - 0.008$. However, for the allowed range of $\varepsilon_e$, the bounds from $\nu-e$ scattering require $|\varepsilon_{\nu}|$ to be reduced by a factor of $\sim 4$ or more.  In this section, we neutralize the $X$-charge of the active neutrinos by supplementing the SM with vectorlike leptons with opposite $B-L$ quantum numbers.  The $B-L$ symmetry is broken by a Higgs mechanism, generating a vacuum expectation value for the new SM-singlet Higgs field $h_X$. This symmetry breaking simultaneously (1) generates the 17~\mev mass for the $X$ boson, (2) generates a Majorana mass for the SM sterile neutrinos, which would otherwise be forbidden by $B-L$ symmetry, and (3) mixes the SM active neutrinos with the new lepton states such that the resulting mass eigenstates have suppressed $X$-charge.   

The fields of these models include the SM Higgs boson $h_{\text{SM}}$, and the SM lepton fields $\ell_L$, $e_R$, and $\nu_R$, where the last is the sterile neutrino required by $B-L$ anomaly cancellation.  To these, we add the Higgs field $h_X$ with $B-L = 2$, and $N$ vectorlike lepton isodoublets $L_{i_{L,R}}$ and charged isosinglets $E_{i_{L,R}}$, with $B-L = 1$.  The addition of vectorlike pairs preserves anomaly cancellation.  These fields and their quantum numbers are shown in \tableref{quantumnumbers}.  We focus here on the first generation leptons; the mechanism may be straightforwardly extended to the second and third generations.

\begin{table}[tb]
 \caption{Fields and their quantum numbers in the $B-L$ model with kinetic mixing and neutrinos neutralized by mixing with vectorlike leptons. The SM fields, including the sterile neutrino, are listed above the line.  The new fields, including $N$ generations of vectorlike fields, with $i = 4, \ldots, N+3$, are listed below the line. }
\vspace*{-.1in}
 \label{table:quantumnumbers}
\begin{tabular}[t]{ccccc}
 \toprule 
Field & \ Isospin $I$ \ & \ Hypercharge $Y$ \ & \ $B-L$ \\
\hline
$h_{\text{SM}}$ & $\frac{1}{2}$ & $\frac{1}{2}$ & 0  \\
$\ell_L = \left( \! \! \begin{array}{c} \nu_L \\ e_L \end{array} \! \! \right)$ & $\frac{1}{2}$ & $- \frac{1}{2}$ & $-1$  \\
$e_R$ & 0 & $-1$ & $-1$  \\
$\nu_R$ & 0 & 0 & $-1$  \\
\hline 
$h_X$ & 0 & 0 & 2 \\
$L_{i_L} = \left( \! \! \begin{array}{c} \nu_{i_L} \\ e_{i_L} \end{array} \! \! \right)$ & $\frac{1}{2}$ & $- \frac{1}{2}$ & 1  \\
$L_{i_R} = \left( \! \! \begin{array}{c} \nu_{i_R} \\ e_{i_R} \end{array} \! \! \right)$ & $\frac{1}{2}$ & $- \frac{1}{2}$ & 1  \\
$E_{i_L}$ & 0 & $-1$ & 1  \\
$E_{i_R}$ & 0 & $-1$ & 1  \\
\toprule
 \end{tabular}
\end{table}

With these fields, the full set of gauge-invariant, renormalizable Lagrangian terms that determine the lepton masses are
\begin{eqnarray}
{\cal L} &=& {\cal L}_{\text{SM}} + {\cal L}_{\text{mix}} + {\cal L}_{\text{new}} \\
{\cal L}_{\text{SM}} &=& 
(- y_e h_{\text{SM}} \bar{\ell}_L e_R + y_{\nu} \tilde{h}_{\text{SM}} \bar{\ell}_L \nu_R + \text{h.c.} )
- y_N h_X \bar{\nu}_R^c \nu_R \\
{\cal L}_{\text{mix}} &=&  
- \lambda_L^i h_X \bar{\ell}_L L_{i_R} 
- \lambda_E^i h_X \bar{E}_{i_L} e_R + \text{h.c.} \label{eq:Lmix} \\
{\cal L}_{\text{new}} &=& 
- M_{L}^{ij} \bar{L}_{i_L} L_{j_R} 
- M_{E}^{ij} \bar{E}_{i_L} E_{j_R}
- h^{ij} h_{\text{SM}} \bar{L}_{i_L} E_{j_R} 
+ k^{ij} \tilde{h}_{\text{SM}} \bar{E}_{i_L} L_{j_R} + \text{h.c.}  \ ,
\label{Lagrangian}
\end{eqnarray}
where $i,j = 4, \ldots, N+3$.  ${\cal L}_{\text{SM}}$ generates the Dirac and Majorana SM neutrino masses, ${\cal L}_{\text{mix}}$ includes the terms that mix the SM and vectorlike fields, and ${\cal L}_{\text{new}}$ contains the vectorlike masses and Yukawa couplings for the new vectorlike leptons.  For simplicity, we will assume universal masses and Yukawa couplings, so $\lambda_{L}^i = \lambda_{L}$, $\lambda_{E}^i = \lambda_{E}$, $M_{L}^{ij} = M_{L} \delta^{ij}$, $M_E^{ij} = M_{E} \delta^{ij}$, $h^{ij} = h \delta^{ij}$, $k^{ij} = k \delta^{ij}$.

When electroweak symmetry and $B-L$ symmetry are broken, the Higgs fields get vevs $\langle h_{\text{SM}} \rangle = v / \sqrt{2}$, where $v \simeq 246~\gev$, and $\langle h_X \rangle = v_X / \sqrt{2}$.  This gives the $X$ boson a mass 
\begin{equation}
m_X = 2 e \, | \varepsilon_{B-L} | \, v_X \ ,
\label{Xmass}
\end{equation}
which constrains $v_X$ to be 
\begin{equation}
v_X = 14~\gev \ \frac{0.002}{| \varepsilon_{B-L} |} \ .
\label{eq:Xvev}
\end{equation}
It also generates Dirac and Majorana masses for the SM neutrinos, $m_D = y_{\nu} v / \sqrt{2}$ and $m_M = y_N v_X/ \sqrt{2}$, and masses $M_I^{L} = \lambda_{L} v_X / \sqrt{2}$ and $M_I^E = \lambda_{E} v_X / \sqrt{2}$  that mix the SM and vectorlike leptons.  The resulting neutrino masses for the first SM generation and the vectorlike generations are $\bar{\psi}^{\nu} {\cal M}_{\nu}^M \psi^{\nu}$, where
\begin{equation}
{\cal M}_{\nu}^M =
\left( \begin{array}{ccccccc}
0 & m_D & 0 & M_I^{L} & \cdots & 0 & M_I^{L} \\
m_D & m_M & 0 & 0 & \cdots & 0 & 0 \\
0 & 0 & 0 & M_{L} & \cdots & 0 & 0 \\
M_I^{L} & 0 & M_{L} & 0  & \cdots & 0 & 0 \\
\vdots & \vdots & \vdots & \vdots & \ddots & \vdots & \vdots \\
0 & 0 & 0 & 0 & \cdots & 0 & M_{L} \\
M_I^{L} & 0 & 0 & 0 & \cdots & M_{L} & 0 
\end{array} \right) \ ,
\label{eq:neutrinomasses}
\end{equation}
and $\psi^{\nu} = (\nu_L , \nu_R, \nu_{4_L}, \nu_{4_R}, \ldots, \nu_{{N+3}_L}, \nu_{{N+3}_R})$, or alternatively, neglecting the small SM Dirac and Majorana masses, the remaining neutrino masses may be written $\bar{\psi}^{\nu}_L {\cal M}_{\nu} \psi^{\nu}_R + \text{h.c.}$, where \begin{equation}
{\cal M}_{\nu} 
= \left( \begin{array}{cccccc}
0 & M_I^{L} & \cdots & M_I^{L} \\
0 & M_{L} & \cdots & 0 \\
\vdots & \vdots & \ddots & \vdots \\
0 & 0 & \cdots & M_{L} 
\end{array} \right) \ ,
\end{equation}
and $\psi^{\nu}_{L,R} = (\nu_{L,R}, \nu_{4_{L,R}}, \ldots, \nu_{{N+3}_{L,R}})$.  Similarly, the charged lepton masses are $\bar{\psi}^{e}_L {\cal M}_{e} \psi^{e}_R + \text{h.c.}$, where
\begin{equation}
{\cal M}_{e} 
= \left( \begin{array}{cccccc}
0 & M_I^{L} & 0 & \cdots & M_I^{L} & 0 \\
0 & M_{L} & \frac{hv}{\sqrt{2}} & \cdots & 0 & 0 \\
M_I^E & \frac{kv}{\sqrt{2}} & M_E & \cdots & 0 & 0 \\
\vdots & \vdots & \vdots & \ddots & \vdots & \vdots \\
0 & 0 & 0 & \cdots & M_{L} & \frac{hv}{\sqrt{2}} \\
M_I^E & 0 & 0 & \cdots & \frac{kv}{\sqrt{2}} & M_E 
\end{array} \right) \ ,
\end{equation}
and $\psi^{e}_{L,R} = (e_{L,R}, e_{4_{L,R}}, \ldots, e_{{N+3}_{L,R}})$.  

Diagonalizing the neutrino mass matrix of \eqref{neutrinomasses} yields $N$ Dirac neutrino states with mass $\sim M_{L}$, and two light states: the SM sterile neutrino and the SM active neutrino, which is the eigenstate 
\begin{equation}
\frac{1}{\sqrt{M_{L}^2 + N M_I^{L\, 2}}} ( - M_{L} , 0, M_I^{L}, 0, M_I^{L} , \ldots , 0 , M_I^{L}, 0) \ .
\end{equation}
The active neutrino's $X$-charge is therefore modified by the mixing with the vectorlike lepton states, with similar effects for the charged leptons.  In the end, we find that the lepton $X$-charges are modified to 
\begin{eqnarray}
\varepsilon_{\nu_L} &=& - \varepsilon_{B-L} \cos 2 \theta_{\nu_L} \\
\varepsilon_{e_L} &=& - \varepsilon_{B-L} \cos 2 \theta_{e_L} - \varepsilon  
= \varepsilon_{B-L} (1 - \cos 2 \theta_{e_L}) -\delta \\
\varepsilon_{e_R} &=& - \varepsilon_{B-L} \cos 2 \theta_{e_R} - \varepsilon 
= \varepsilon_{B-L} (1 - \cos 2 \theta_{e_R}) -\delta \ , 
\end{eqnarray}
where
\begin{equation}
\tan \theta_{\nu_L} = \frac{N M_I^{L\, 2}}{M_{L}^2} \ ,
\end{equation}
and $\theta_{e_L}$ and $\theta_{e_R}$ are determined by similar, but more complicated, relations derived by diagonalizing ${\cal M}_e$.  To neutralize the neutrino charge, we need
\begin{equation}
\tan \theta_{\nu_L} = \frac{N M_I^{L\, 2}}{M_{L}^2 } 
= \frac {N \lambda_L^2 m_X^2}  {8 M_{L}^2 e^2 \varepsilon_{B-L}^2}
\approx  \biggl[ \frac{130~\gev}{M_{L}} \biggr]^2
\biggl[ \frac{0.002}{\varepsilon_{B-L}} \biggr]^2
\biggl[ \frac{\sqrt{N} \lambda_{L}}{4 \pi} \biggr]^2 \approx 1 \ ,
\label{neutralization}
\end{equation}
where we have normalized the effective coupling $\sqrt{N} \lambda_L$ to its ultimate perturbative limit. We see that the neutrino $X$-charge may be neutralized with as few as $N=1$ vectorlike lepton generation with mass at the weak scale.  A larger number of heavier vectorlike leptons may also neutralize the neutrino $X$-charge. In addition, to preserve non-chiral electron couplings, we require $\theta_{e_L} \approx \theta_{e_R}$.

\subsection{Implications for Colliders and Cosmology}
\label{sec:activeimplications}

Here we consider the implications of these models for colliders and cosmology, beginning with the extremely simple case of $N=1$ generation of vectorlike leptons and vanishing Yukawa couplings $h = k = 0$.  In this case, the mass matrices are easily diagonalized.  The heavy states include three ``4th generation'' Dirac fermions: the isodoublet neutrino and electron with masses $m_{\nu_4} \simeq m_{e_4} \simeq \sqrt{2} M_{L}$ and the isosinglet electron with mass $m_{E_4} \simeq \sqrt{2} M_E$.  The states $\nu_4$ and $e_4$ have vectorlike masses and are nearly degenerate, and so do not contribute to the $S$ and $T$ parameters~\cite{Peskin:1991sw}.  The light states are the usual massless SM leptons, but mixed with opposite $X$-charged states, with mixing angles $\tan \theta_{\nu_L} = \tan \theta_{e_L} = (M_I^{L} / M_{L})^2$ and $\tan \theta_{e_R} = (M_I^{E} / M_E)^2$.  These SM fields each mix only with new leptons with the same SM quantum numbers, and so these mixing angles are not constrained by precision measurements.  Choosing $M_I^{L} / M_{L} = M_I^{E} / M_E = 1$, we find $\varepsilon_{\nu_L} = 0$ and $\varepsilon_{e_L} = \varepsilon_{e_R} = \varepsilon_{B-L} -\delta$.  For $\varepsilon_{B-L} \approx 0.002$ and $\delta \approx 0.001$, this extension of the SM contains a protophobic gauge boson that explains the $^8$Be signal consistent with all current constraints. As in the U(1)$_B$ case, assuming $\varepsilon_{\mu} \approx \varepsilon_e$ removes at least part of the $(g-2)_{\mu}$ puzzle and implies promising prospects for future searches at low-energy colliders, as discussed in \secref{conclusions}.

The new vectorlike leptons can be produced through Drell-Yan production at hadron and $e^+ e^-$ colliders, and so this model may be explored at the LHC and future colliders. The prospects for vectorlike lepton searches at the LHC have been studied in detail in the case that they decay to $W \nu_{\ell}$, $Z \ell$, and $h \ell$~\cite{Dermisek:2014qca,Kumar:2015tna,Abdullah:2016avr}.  In the present case, however, the vectorlike lepton masses and decays are constrained by the neutrino neutralization mechanism.  In particular,  the mixing terms of \eqref{Lmix} that neutralize the neutrinos imply that the decays $\nu_4 \to \nu_e h_X$, $e_4 \to e h_X$, and $E_4 \to e h_X$ are almost certainly dominant.  

The $B-L$ Higgs boson has a variety of possible decays, but for a moderately large Majorana Yukawa coupling $y_N$, the invisible decay $h_X \to \nu_R \bar{\nu}_R$ dominates.  The resulting processes are therefore 
\begin{alignat}{3}
pp 
&\to\; E_4^+ E_4^- 
&&\to\; e^+ e^- h_X h_X 
&&\to\; e^+ e^- \nu_R \bar{\nu}_R \nu_R \bar{\nu}_R 
\\
pp 
&\to\; e_4^+ e_4^- 
&&\to\; e^+ e^- h_X h_X 
&&\to\; e^+ e^- \nu_R \bar{\nu}_R \nu_R \bar{\nu}_R 
\\
pp 
&\to\; \nu_4 \bar{\nu}_4 
&&\to\; \nu_L \bar{\nu}_L h_X h_X 
&&\to\; \nu_L \bar{\nu}_L \nu_R \bar{\nu}_R \nu_R \bar{\nu}_R 
\\
pp 
&\to\; \nu_4 e_4 
&&\to\; \nu e h_X h_X 
&&\to\; e \nu_L \nu_R \bar{\nu}_R \nu_R \bar{\nu}_R  \ .
\end{alignat}
These signals are therefore very similar to those of selectron pair production and selectron--sneutrino pair production, leading to signatures with missing transverse energy $\mET$, $e^+ e^- + \mET$ and $e^{\pm} + \mET$. 
The amount of missing energy is controlled by 
\begin{equation}
m_{h_X} = \sqrt{\lambda_H} v_X = 
70~\gev \, \sqrt{\frac{\lambda_H}{4\pi}} \, \frac{0.002}{| \varepsilon_{B-L} |} \ ,
\end{equation}
where $\lambda_H$ is the Higgs boson quartic coupling appearing in the Lagrangian term $\lambda_H (h_X h_X^*)^2$, and we have used \eqref{Xvev}.    

Current bounds from the combination of LEP2 and 8~TeV LHC data on the combined production of right- and left-handed selectron and smuons with mass 100 GeV allow neutralino masses of around 50 GeV~\cite{Aad:2014vma,Khachatryan:2014qwa}.  The vectorlike lepton cross section is bigger by roughly a factor of 4, but 100 GeV vectorlike leptons decaying to $50-70~\gev$ $B-L$ Higgs bosons may still be allowed. Existing mono-lepton searches based on 8~TeV LHC data are not optimized for lepton masses as low as 100 GeV and are unlikely to have sensitivity~\cite{Khachatryan:2014tva,ATLAS:2014wra}.  Nonetheless, it may be that future searches based on 13 ~TeV data will become sensitive, particularly if they can be optimized for lower mass vectorlike leptons. It would be interesting to investigate this scenario in more detail, as well as scenarios where other $h_X$ decays are comparable or dominant to the invisible decay assumed above.  It is also worth noting that the appearance of relatively strong couplings ($\lambda_H$, $\lambda_L$) in the $h_X$ sector may be an indication of compositeness, which could result in a richer and more complicated set of final states accessible to LHC energies.

We now turn to the SM neutrino sector and potential cosmological signatures. As noted above, when the $h_X$ field with $B-L$ charge 2 gets a vev, it also generates a Majorana mass for the SM singlet neutrinos.  This is an important feature.  Without a charge 2 Higgs boson, the SM neutrinos are Dirac particles.  Light Dirac neutrinos are not typically problematic, as the $\nu_R$ component does not thermalize and does not contribute to the number of relativistic degrees of freedom $n_{\text{eff}}$.  In the current model, however, the process $f \bar{f} \leftrightarrow X \leftrightarrow \bar{\nu}_R \nu_R$ effectively thermalizes the $\nu_R$ at temperatures $T \sim m_X$, where the process is on-resonance.  To avoid thermalization, one needs the $X$-charge of $\nu_R$ to be less than $10^{-9}$~\cite{Heeck:2014zfa} or very low reheat temperatures in the window between 1 MeV and $m_X \approx 17~\mev$.  The generation of a Majorana mass avoids these problems.  

The Majorana mass is 
\begin{equation}
m_M = y_N v_X / \sqrt{2} = y_N \frac{m_X}{2 \sqrt{2} e |\varepsilon_{B-L}| } \alt 30~\gev \,
\frac{0.002}{|\varepsilon_{B-L}| }\ ,
\end{equation}
where the upper bound assumes $y_N \sim 3$.  The physical masses of the SM active neutrinos are then determined by the see-saw mechanism, with Dirac masses chosen appropriately.  Of course, the sterile neutrino masses need not be near their upper limit, and it is tempting to postulate that they may be in the keV range as required for warm dark matter.  To prevent the decays $X \to \nu_R \nu_R$ from significantly diluting the $^8$Be signal in this case, the $\nu_R$ $X$-charges must also be neutralized, for example, through mixing with vectorlike isosinglet neutrinos.  Alternatively, the sterile neutrino masses may be in the 10 -- 100~\mev range, as may be helpful for reducing the standard BBN predictions for the $^7$Li abundance to the observed levels~\cite{Goudelis:2015wpa}.  We leave these astrophysical and cosmological implications for future work.

One might worry that having a model with an exact U(1)$_{B-L}$ or U(1)$_{B}$ gauge symmetry down to the GeV or MeV energy scale would prevent any baryon number asymmetry from being generated. This, however, is not the case, as was discussed, for example, in Ref.~\cite{Duerr:2014wra} for a model with gauged U(1)$_B$. A lepton number asymmetry can still be produced at a high scale and then be partially converted into baryon number through the electroweak sphalerons. For the case of gauged U(1)$_{B-L}$ one could also invoke a Dirac leptogenesis scenario which relies on the fact that the right-handed neutrinos decouple early on during the evolution of the Universe, trapping some amount of lepton number~\cite{Dick:1999je,Murayama:2002je}. The resulting lepton number deficit in the visible sector is then again transferred to baryon number through the sphalerons.

We have introduced additional fermionic matter to render the models 
compatible with experimental constraints. The step of adding extra matter may not be necessary, and it may be possible to satisfy all the existing experimental constraints by considering a combination of gauged U(1) quantum numbers. The possibility of multiple, new U(1) gauge bosons has been explored previously, in the two dark-photon (``paraphoton'') case~\cite{Masso:2006gc} and for three Abelian groups~\cite{Heeck:2011md}.  Here we note that if one were to combine a U(1)$_{B-L}$ model with kinetic mixing with a second, unbroken (or softly broken) 
gauge symmetry, e.g., $L_e - L_\tau$, it is possible to bring the first-generation 
fermion couplings of the $B-L$ gauge boson to the form of the U(1)$_B$ model. Such relationships are completely compatible with the couplings needed to describe the $^8$Be anomaly and satisfy other constraints. However, equivalence principle constraints on new, massless gauge bosons that can couple to the constitutents of ordinary matter are severe~\cite{Schlamminger:2007ht,Wagner:2012ui,Heeck:2014zfa}. We note that we can address this problem by making the massless gauge boson's couplings to electrons vanish at tree level. Further investigation is required to check that this suffices to render the model compatible with experimental constraints on new, (nearly) massless gauge bosons.

\section{Future Experiments}
\label{sec:future:experiments}

\begin{figure}[tb]
\includegraphics[width=.5\linewidth]{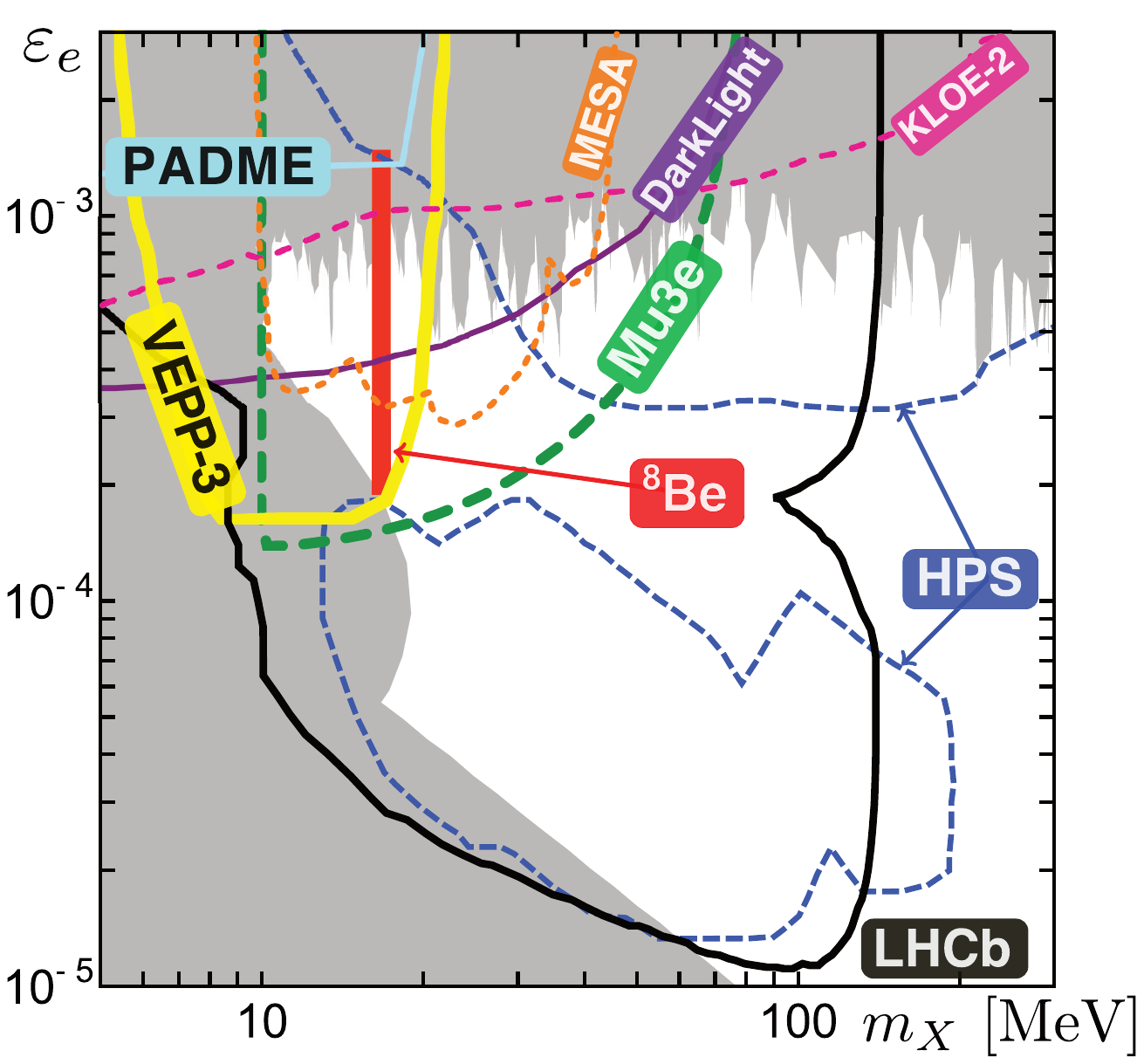} 
\vspace*{-0.1in}
\caption{The $^8$Be signal region, along with current constraints (gray) and projected sensitivities of future experiments in the $(m_X, \varepsilon_e)$ plane.  
	Updated from Ref.~\cite{Feng:2016jff}. Note ${\rm Br}(X\to e^+e^-)=1$ is assumed. 
\label{fig:future} }
\end{figure}

Current and near future experiments will probe the parameter space of interest for the protophobic gauge boson $X$. The projected sensitivities of various experiments are shown in \figref{future} and we briefly discuss them below. 

\paraflip{Other Large Energy Nuclear Transitions.}  The \bestar and \bestarprime states are quite special in that they decay electromagnetically to discrete final states with an energy release in excess of 17~\mev. Other large-energy gamma transitions have been observed~\cite{Robinson}, such as the 19.3~\mev transition in $^{10}$B to its ground state~\cite{Ling:1971oik} and the 17.79~\mev transition in $^{10}$Be to its ground state~\cite{Subotic:1978mab}.  Of course, what is required is large production cross sections and branching fractions so that many IPC events can be observed.  It would certainly be interesting to identify other large energy nuclear transitions with these properties to test the new particle interpretation of the $^8$Be anomaly.  

\paraflip{LHCb.}
A search for dark photons $A'$ at LHCb experiment during Run 3 (scheduled for the years 2021 -- 2023) has been proposed~\cite{Ilten:2015hya} using the charm meson decay $D^*(2007)^0 \to D^0 A'$ with subsequent $A' \to e^+ e^-$. It takes advantage of the LHCb excellent vertex and  invariant mass resolution. For dark photon masses below about 100 MeV, the experiment  can explore nearly all of the remaining parameter space in $\varepsilon_e$ between the existing prompt-$A'$ and beam-dump limits. In particular, it can probe the entire region relevant for the $X$ gauge boson explaining the $^8$Be anomaly.

\paraflip{Mu3e.}
The Mu3e experiment will look at the muon decay channel $\mu^+ \to e^+ \nu_e \bar{\nu}_\mu (A'\to e^+ e^-)$ and will be sensitive to dark photon masses in the range $10 ~\mev \lesssim m_{A'} \lesssim 80 ~\mev$~\cite{Echenard:2014lma}. The first phase (2015 -- 2016) will probe the region $\varepsilon_e \gtrsim 4\times 10^{-3}$, while phase II (2018 and beyond) will extend this reach almost down to $\varepsilon_e \sim 10^{-4}$, which will include the whole region of interest for the protophobic gauge boson $X$. 

\paraflip{VEPP-3.}
A proposal for a new gauge boson search at the VEPP-3 facility was made~\cite{Wojtsekhowski:2012zq}. The experiment will consist of a positron beam incident on a gas hydrogen target and will look for missing mass spectra in $e^+ e^- \to A' \gamma$. The search will be independent of the $A'$ decay modes and lifetime. Its region of sensitivity in $\varepsilon_e$ extends down into the beam dump bounds, i.e., below $\varepsilon_e \sim 2\times 10^{-4}$, and includes the entire region relevant for $X$. Once accepted, the experiment will take 3 -- 4 years.

\paraflip{KLOE-2.}
As mentioned above, the KLOE-2 experiment, looking for $e^+ e^- \to \gamma (X \to e^+ e^-)$, is running and improving its current bound of  $| \varepsilon_e | < 2 \times 10^{-3}$~\cite{Anastasi:2015qla} for $m_X \approx 17~\mev$. With the increased DA$\phi$NE-2 delivered luminosity and the new detectors, KLOE-2 is expected to improve this limit by a factor of two within two years~\cite{Graziani2016}.  

\paraflip{MESA.}
The MESA experiment will use an electron beam incident on a gaseous target to produce dark photons of masses between $\sim 10 \!- \!40 ~\mev$ with electron coupling as low as $\varepsilon_e \sim 3 \times 10^{-4}$, which would probe most of the available $X$ boson parameter space~\cite{Beranek:2013yqa}. The commissioning is scheduled for 2020.

\paraflip{DarkLight.} 
The DarkLight experiment, similarly to VEPP-3 and MESA, will use electrons scattering off a gas hydrogen target to produce on-shell dark photons, which later decay to $e^+e^-$ pairs~\cite{Balewski:2014pxa}. It is sensitive to masses in the range  $10\!-\!100 ~\mev$ and $\varepsilon_e$ down to $4 \times 10^{-4}$, covering the majority of the allowed protophobic $X$ parameter space. Phase I of the experiment is expected to take data in the next 18 months, whereas phase II could run within two years after phase I. 

\paraflip{HPS.}
The Heavy Photon Search experiment is using a high-luminosity electron beam incident on a tungsten target to produce dark photons and search for both $A'\to e^+ e^-$ and $A'\to \mu^+ \mu^-$ decays~\cite{Moreno:2013mja}. Its region of sensitivity is split into two disconnected pieces (see \figref{future}) based on the analyses used: the upper region is probed solely by a bump hunt search, whereas the lower region also includes a displaced vertex search. HPS is expected to complete its dataset by 2020.

\paraflip{PADME.}
The PADME experiment will look for new light gauge bosons resonantly produced in collisions of a positron beam with a diamond target, mainly through the process $e^+e^- \to X \gamma$~\cite{Raggi:2015gza}. The collaboration aims to complete the detector assembly by the end of 2017 and accumulate $10^{13}$ positrons on target by the end of 2018. The expected sensitivity after one year of running is $\varepsilon_e \sim 10^{-3}$, with plans to get as low as $10^{-4}$~\cite{Fayet:2007ua,Raggi:2014zpa}.

\paraflip{BES III.} Current and future $e^+ e^-$ colliders, may also search for $e^+e^- \to X \gamma$. A recent study has explored the possibility of using BES III and BaBar to probe the 17~\mev protophobic gauge boson~\cite{Chen:2016dhm}.

\paraflip{E36 at J-PARC (TREK).} 
The TREK experiment has the capacity to study $K\to \mu \nu e^+e^-$ decays~\cite{Kohl:2013rma}; 
the enhancement associated with the interaction of the longitudinal component of $X$ with 
charged fermions 
should make for sensitive tests of $\varepsilon_e$ in the mass range of interest to the 
$^8$Be anomaly~\cite{Carlson:2013mya}.

\section{Conclusions}
\label{sec:conclusions}

The 6.8$\sigma$ anomaly in $^8$Be cannot be plausibly explained as a statistical fluctuation, and the fit to a new particle interpretation has a $\chi^2$/dof of 1.07. If the observed bump has a nuclear physics or experimental explanation, the near-perfect fit of the $\theta$ and $m_{ee}$ distributions to the new particle interpretation is a remarkable coincidence.  Clearly all possible explanations should be pursued.  Building on our previous work~\cite{Feng:2016jff}, in this study, we presented particle physics models that extend the SM to include a protophobic gauge boson that explains the $^8$Be observations and is consistent with all other experimental constraints.  

To understand what particle properties are required to explain the $^8$Be anomaly, we first presented effective operators for various spin-parity assignments.  Many common examples of light, weakly coupled particles, including dark photons, dark Higgs bosons, axions, and $B-L$ gauge bosons (without kinetic mixing) are disfavored or excluded on general grounds.  In contrast, general gauge bosons emerge as viable candidates.

In Ref.~\cite{Feng:2016jff} we determined the required couplings of a vector gauge boson to explain the $^8$Be anomaly assuming isospin conservation, and found that the particle must be protophobic.  In this work, we refined this analysis to include the possibility of isospin mixing in the \bestar and \bestarprime states.  Although isospin mixing and violation can yield drastically different results, these effects are relatively mild once one focuses on protophobic gauge bosons. It would be helpful to have a better understanding of the role of isospin breaking in these systems and a quantitative estimate of their uncertainties. The presence of isospin mixing also implies that the absence of an anomaly in \bestarprime decays must almost certainly be due to kinematic suppression and that the $X$ particle's mass is above 16.7 MeV. Combining all of these observations with constraints from other experiments, we then determined the favored couplings for any viable vector boson explanation. 

 We have presented two anomaly-free extensions of the SM that resolve the $^8$Be anomaly. In the first, the protophobic gauge boson is a U(1)$_B$ gauge boson that kinetically mixes with the photon.  For gauge couplings and kinetic mixing parameters that are comparable in size and opposite in sign, the gauge boson couples to SM fermions with approximate charge $Q-B$, satisfying the protophobic requirement.  Additional matter content is required to cancel gauge anomalies, and we presented a minimal set of fields that satisfy this requirement.  In the second model, the gauge boson is a U(1)$_{B-L}$ gauge boson with kinetic mixing, and the SM fermion charges are $Q-(B-L)$. Additional vectorlike leptons are needed to neutralize the neutrino if we consider only a single U(1) gauge group.  Both models can simultaneously resolve the $(g-2)_{\mu}$ anomaly, have large electron couplings that can be probed at many near future experiments, and include new vectorlike lepton states at the weak scale that can be discovered by the LHC. 

One may speculate that the protophobic gauge boson may simultaneously resolve not only the $^8$Be and $(g-2)_{\mu}$ anomalies, but also others.  Possibilities include the NuTeV anomaly~\cite{Liang:2016ffe} and the cosmological lithium problem mentioned in \secref{activeimplications}. Another possibility is the $\pi^0 \to e^+e^-$ KTeV anomaly, which may be explained by a spin-1 particle with axial couplings that satisfy
\begin{align}
	\left(g_A^u - g_A^d\right) g_A^e \left(\frac{20~\mev}{m_X}\right)^2 \approx 1.6 \times 10^{-7}\, ,
\end{align}
which is roughly consistent with the vector couplings we found for a protophobic gauge boson~\cite{Kahn:2007ru}. Independent of experimental anomalies, a spin-1 boson with purely axial couplings is a promising candidate for future study \cite{Kahn:2016vjr}. Such bosons need not be protophobic, because their suppressed contributions to neutral pion decays relax many constraints that existed for vector bosons. We note, however, that some bounds become stronger for the axial case. For example, the decay $\phi\to \eta (X \to e^+e^-)$ used in deriving the KLOE constraints~\cite{Babusci:2012cr} is an $s$-wave in the axial case, implying a stronger bound than the $p$-wave--suppressed one in the vector case.  Another example is $(g-2)_e$~\cite{Aoyama:2012wj}, for which an axial vector makes larger contributions than a vector, for couplings of the same magnitude.  In addition, there are very stringent bounds, for example, from atomic parity violation, on gauge bosons with mixed vector and axial vector couplings~\cite{Davoudiasl:2012ag}.   

Finally, if the $^8$Be anomaly is pointing toward a new gauge boson and force, it is natural to consider whether this force may be unified with the others, with or without supersymmetry.  In the case of U(1)$_{B-L}$, which is a factor of many well-motivated grand unified groups, it is tempting to see whether the immediately obvious problems---for example, the hierarchy between the required U(1)$_{B-L}$ gauge coupling and those of the SM---can be overcome, and whether MeV-scale data may be telling us something interesting about energy scales near the Planck scale.

\acknowledgments 

We thank Phil Barbeau, John Beacom, Roy Holt, Yoni Kahn, Attila J.~Krasznahorkay, Saori Pastore, Tilman Plehn, Mauro Raggi, Alan Robinson, Martin Savage, Paolo Valente, and Robert Wiringa for helpful correspondence.  J.L.F., B.F., I.G., J.S., T.M.P.T., and  P.T.\ are supported in part by  NSF Grants PHY-1316792 and PHY-1620638. The work of S.G.\ is supported in part by the DOE Office of Nuclear Physics under contract DE-FG02-96ER40989.  The work of J.L.F.\ is supported in part by a Guggenheim Foundation grant and in part by Simons Investigator Award \#376204 and was performed in part at the Aspen Center for Physics, which is supported by NSF Grant No.~PHY-1066293.

\vspace*{.2in}

\noindent Note added: Following this paper, the pseudoscalar and axial vector cases have been further studied in Refs.~\cite{Ellwanger:2016wfe} and \cite{Kozaczuk:2016nma}, respectively.

\providecommand{\href}[2]{#2}\begingroup\raggedright\endgroup

\end{document}